\documentclass[12pt,reqno]{amsart}
\usepackage{amssymb, amsmath, hyperref}
\usepackage{pdfsync}
\usepackage{stmaryrd}
\usepackage{bbold}
\usepackage{bbm}
\usepackage{graphics,graphicx,fancyhdr,color}

\newtheorem{df}{Definition}[section]
\newtheorem{lm}[df]{Lemma}
\newtheorem{lemma}[df]{Lemma}
\newtheorem{thm}[df]{Theorem}

\newtheorem{condition}[df]{Condition}

\makeatletter \@addtoreset{equation}{section}

\newcommand{\cal}{\mathcal}

\def\i{{\rm i}}

\newcommand{\bes}{\begin{displaymath}}
\newcommand{\ees}{\end{displaymath}}
\newcommand{\be}{\begin{equation}}
\newcommand{\ee}{\end{equation}}
\newcommand{\ba}{\begin{eqnarray}}
\newcommand{\ea}{\end{eqnarray}}
\newcommand{\bas}{\begin{eqnarray*}}
\newcommand{\eas}{\end{eqnarray*}}
\newcommand{\@Bbb}[1]{\ensuremath{\mathbb #1}}

\newcommand{\B}{{\@Bbb B}}
\newcommand{\C}{{\@Bbb C}}
\newcommand{\E}{{\mathbb E}}

\newcommand{\F}{{\@Bbb F}}
\renewcommand{\P}{{\mathbb P}}
\newcommand{\bbP}{{\P}}
\newcommand{\bbE}{{\mathbb E}}
\newcommand{\Q}{{\@Bbb Q}}
\newcommand{\bQ}{{\@Bbb Q}}
\newcommand{\N}{{\@Bbb N}}
\newcommand{\R}{{\mathbb R}}
\newcommand{\1}{{\mathbb 1}}
\newcommand{\bbR}{{\@Bbb R}}
\newcommand{\W}{{\@Bbb W}}
\newcommand{\Z}{{\mathbb Z}}
\newcommand{\bbZ}{{\@Bbb Z}}
\newcommand{\bbT}{{\@Bbb T}}

\newcommand{\ls}{\left[}
\newcommand{\rs}{\right]}
\newcommand{\lc}{\left(}
\newcommand{\rc}{\right)}
\newcommand{\labs}{\left|}
\newcommand{\rabs}{\right|}
\newcommand{\lcu}{\left\{}
\newcommand{\rcu}{\right\}}

\newcommand{\la}{\lambda}
\newcommand{\al}{\alpha}
\newcommand{\ka}{\kappa}

\newcommand{\si}{\sigma}

\newcommand{\Om}{\Omega}
\newcommand{\om}{\omega}

\newcommand{\eps}{\epsilon}

\newcommand{\@s}[1]{\ensuremath{\mathcal #1}}
\newcommand{\cA}{\@s A}
\newcommand{\cB}{\@s B}
\newcommand{\cC}{\@s C}
\newcommand{\cD}{\@s D}
\newcommand{\cE}{\@s E}
\newcommand{\cF}{\@s F}
\newcommand{\cG}{\@s G}
\newcommand{\cH}{\@s H}
\newcommand{\cI}{\@s I}
\newcommand{\cJ}{\@s J}

\newcommand{\cK}{\@s K}
\newcommand{\cL}{\@s L}
\newcommand{\cN}{\@s N}
\newcommand{\cM}{\@s M}
\newcommand{\cO}{\@s O}
\newcommand{\cP}{\@s P}
\newcommand{\cR}{\@s R}
\newcommand{\cS}{\@s S}
\newcommand{\cT}{\@s T}
\newcommand{\cV}{\@s V}
\newcommand{\cW}{\@s W}
\newcommand{\cX}{\@s X}
\newcommand{\cY}{\@s Y}
\newcommand{\cZ}{\@s Z}

\newcommand{\@bm}[1]{\ensuremath{\mathbf #1}}
\newcommand{\bma}{\@bm a}
\newcommand{\bmb}{\@bm b}
\newcommand{\bmc}{\@bm c}
\newcommand{\bmd}{\@bm d}

\newcommand{\bme}{\@bm e}
\newcommand{\bmf}{\@bm f}
\newcommand{\bmg}{\@bm g}
\newcommand{\bmh}{\@bm h}
\newcommand{\bmi}{\@bm i}
\newcommand{\bmj}{\@bm j}
\newcommand{\bmk}{\@bm k}
\newcommand{\bml}{\@bm l}
\newcommand{\bmm}{\@bm m}
\newcommand{\bmn}{\@bm n}
\newcommand{\bmo}{\@bm o}
\newcommand{\bmp}{\@bm p}
\newcommand{\bmq}{\@bm q}
\newcommand{\bmr}{\@bm r}
\newcommand{\bms}{\@bm s}
\newcommand{\bmt}{\@bm t}
\newcommand{\bmu}{\@bm u}
\newcommand{\bmw}{\@bm w}
\newcommand{\bmv}{\@bm v}
\newcommand{\bmx}{\@bm x}
\newcommand{\bx}{\@bm x}

\newcommand{\bmy}{\@bm y}
\newcommand{\bz}{\@bm z}

\newcommand{\by}{\@bm y}

\newcommand{\bmzero}{\@bm 0}

\newcommand{\ga}{\gamma}

\newcommand{\@g}[1]{\ensuremath{\mathfrak #1}}
\newcommand{\gA}{\@g A}
\newcommand{\gD}{\@g D}
\newcommand{\gJ}{\@g J}
\newcommand{\gF}{\@g F}
\newcommand{\gM}{\@g M}
\newcommand{\gR}{\@g R}

\newcommand{\commentout}[1]{{}}

\makeatother

\begin{document}

\title[Convergence rates]{Long time, large scale limit of the Wigner transform for a system of linear oscillators  in one dimension}

\author{Tomasz Komorowski}
\author{\L ukasz St\c{e}pie\'n}
\begin{abstract}
We consider the long time, large scale behavior of
the Wigner transform $W_\eps(t,x,k)$ of  the wave function corresponding to  a discrete wave equation on a  $1$-d  integer lattice, with a weak multiplicative noise. 
This model has been introduced in \cite{BBO1} to describe a   system of interacting linear oscillators with a weak noise that conserves locally the kinetic energy and the momentum. 
The kinetic limit for the Wigner transform has been shown in \cite{BOS}. In the present paper we prove that   in the unpinned case   there exists $\gamma_0>0$ such that  for any $\gamma\in(0,\gamma_0]$ the weak limit of $W_\eps(t/\eps^{3/2\gamma},x/\eps^{\gamma},k)$, as $\eps\ll1$, satisfies a one dimensional fractional heat equation $\partial_t W(t,x)=-\hat c(-\partial_x^2)^{3/4}W(t,x)$ with $\hat c>0$.  In the pinned case an analogous result can be claimed for $W_\eps(t/\eps^{2\gamma},x/\eps^{\gamma},k)$ but the limit satisfies then the usual heat equation.
\end{abstract}

\vspace*{-1in}

\maketitle

\section{Introduction}

%
%
%

In the present paper we are concerned with the asymptotic behavior of the Wigner transform of the wave function corresponding to   a discrete wave equation  on a one dimensional integer lattice with a weak multiplicative noise, see \eqref{basic:sde} below. This kind of an equation arises naturally while considering  a stochastically perturbed   chain   of  oscillators with harmonic interactions, see \cite{BBO1} and also \cite{luspohn}. It has been argued in \cite{BBO1}  that, due to the presence of the noise conserving both the energy and the momentum (in fact the latter property is crucial), in the low dimensions ($d=1$, or $2$)   conductivity of  this explicitly solvable model diverges as $N^{1/2}$ in dimension $d=1$ and $\log N$, when $d=2$, where $N$ is the length of the chain.
This complies with numerical results concerning  some  anharmonic chains with no noise, see e.g. \cite{llp} and \cite{LLP}. We refer an interested reader to the review papers \cite{BRB,LLP} and the references therein for more background information on the subject of heat transport in anharmonic crystals.

   It has been shown in \cite{BOS} that in the weakly coupled case, i.e. when the coupling parameter $\eps$ is small, the asymptotics of the Wigner function $W_\eps(t,x,k)$ (defined below  by \eqref{wigner-gam}, with $\gamma=0$), that describes the resolution of the energy in spatial and momentum coordinates $(x,k)$ at time $t\sim \eps^{-1}$,  is given by a linear Boltzmann equation, see \eqref{lattice-wave} below. Furthermore, since  in the dimension $d=1$ the scattering rate of a phonon is of order $k^2$ for small wavenumber $k$,  in the unpinned case   (then the dispersion relation satisfies $\om'(k)\approx {\rm sign\, }k$, for $|k|\ll1$) the long time, large space asymptotics of the solution of the transport equation     can be described by a fractional (in space) heat equation 
$$
\partial_tW(t,x)=-\hat c(-\partial^2_x)^{3/4}W(t,x)
$$ for some $\hat c>0$. The initial condition $W(0,x)$ is the limit  of the average of the initial Wigner transform over the wavenumbers, see \cite{jko} and  also \cite{BB,mmm}. Note that the above equation is invariant under time-space scalings $t\sim t'/\eps^{3\gamma/2}$, $x\sim x'/\eps^{\gamma}$ for an arbitrary $\gamma>0$.  This suggests that the fractional heat equation is the limit of the Wigner transform in the above time space scaling.
In our first main result, see part 1) of Theorem \ref{sketch} below, we prove that it is indeed the case when $\gamma\in(0,\gamma_0]$ for some $\gamma_0>0$.

On the other hand, in the pinned
case, i.e. when the dispersion relation satisfies $\om'(k)\approx 0$, as $|k|\ll1$, one can show that the solution of the Boltzmann equation approximates the regular heat equation 
\begin{equation}
\label{heat}
\partial_tW(t,x)=\hat c\partial^2_xW(t,x).
\end{equation}
In part 2) of Theorem \ref{sketch} below we assert  that  if the Wigner transform is considered under the scaling $(t,x)\sim(t'/\eps^{2\ga},x'/\eps^{\ga})$, for $\gamma\in(0,\gamma_0]$ and some $\gamma_0>0$, then it converges to $W(t,x)$, as $\eps\ll1$. 
The coefficient $\hat c$ in the heat equation \eqref{heat} differs from the thermal conductivity coefficient  calculated explicitly  in the pinned case with the help of the Green-Kubo formula  in   \cite{BBO2}, see Theorem 1.
In fact, its computation, see formulas \eqref{031207a} - \eqref{031207b} below, requires solving a Poisson equation \eqref{031207c} and does not lead to an explicit formula. 
Finally, we mention also the results concerning the diffusive limits  for the Wigner transform of a solution of the wave equation on a lattice with a random local velocity in the weak coupling regime (see \cite{luspohn}), for the  geometric optics  regime for the wave equation in continuum  (see \cite{kr}) and in the case of Schr\"odinger equation in the radiative transport regime (see    \cite{esy}).


The proof of Theorem  \ref{sketch} is made of two principal ingredients:   the estimates of the convergence rate for the Wigner transform towards the solution of  the kinetic equation, see Theorem \ref{thm-main1} below in the unpinned case (resp. Theorem \ref{thm-main1a} for the pinned case), and the respective rate of convergence estimates for the solutions of the scaled kinetic equation, see Theorem \ref{prop1} (resp. Theorem \ref{prop1a}). 
 To prove the latter  we  
 show two probabilistic results: Theorems \ref{main-1} and \ref{main-5} that  are of interest on their own. They
provide estimates of  the rate of convergence of the characteristic functions corresponding to a scaled  additive functional of a stationary Markov chain towards the characteristic function of an appropriate stable limit and the respective result in the continuous
time case.

\section{Description of the model and preliminaries}

\subsection{Discrete wave equation with a noise}

We consider a   discrete wave equation with the multiplicative noise on a one dimensional integer lattice, see \cite{BBO1},
\begin{equation}
\label{basic:sde}
\begin{cases} \dfrac{d{\frak q}_x}{dt}=\partial_{{\frak p}_x}{\cal H}( {\frak p}, {\frak q})\\
\\
 \dfrac{d{\frak p}_x}{dt}=-\partial_{{\frak q}_x}{\cal H}({\frak p}, {\frak q})+
\dot\xi_x^{(\eps)}(t).
\end{cases}
\end{equation}
Here $( {\frak p},{\frak q} )=\{( {\frak p}_x,{\frak q}_x),\,x\in\bbZ\}$, where  the  component  labelled by $x$ corresponds to the  one dimensional momentum ${\frak p}_x$ and
position ${\frak q}_x$.
The Hamiltonian corresponds to an infinite  chain of harmonic oscillators and is given by
$$
{\cal H}( {\frak p}, {\frak q}):=\frac12\sum_{y\in\bbZ}{\frak p}_y^2+\frac12\sum_{y,y'\in\bbZ}\alpha(y-y'){\frak q}_y{\frak q}_{y'}.
$$
The interaction potential  $\{\alpha_x,\,x\in\bbZ\}$  will be further specified later on.
 The noises $\{\dot\xi_x^{(\eps)}(t),\,x\in\bbZ\}$ are defined by the following stochastic differentials 
\begin{eqnarray}
\label{noise-strat}
d\xi_{x}^{(\eps)}(t)=\sqrt{\eps}\sum_{k=-1,0,1}(Y_{x+k}{\frak p}_x)\circ dw_{x+k}(t),
\end{eqnarray}
understood  in the Stratonovich sense. Here
$$
Y_x:=({\frak p}_x-{\frak p}_{x+1})\partial_{{\frak p}_{x-1}}+({\frak p}_{x+1}-{\frak p}_{x-1})\partial_{{\frak p}_{x}}+({\frak p}_{x-1}-{\frak p}_{x})\partial_{{\frak p}_{x+1}}
$$
and $\{w_{x}(t),\,t\ge0\}$, $x\in\bbZ$ are i.i.d. standard, one dimensional Brownian motions over a certain probability space $(\Om,{\cal F},\bbP)$. 
Note that the vector field $Y_x$ is tangent to the surfaces  
\begin{equation}
\label{051207a}
{\frak p}_{x-1}^2+{\frak p}_x^2+{\frak p}_{x+1}^2\equiv {\rm const}
\end{equation} 
and 
\begin{equation}
\label{051207b}
{\frak p}_{x-1}+{\frak p}_x+{\frak p}_{x+1}\equiv {\rm const}
\end{equation} 
therefore the system \eqref{basic:sde} conserves the total energy and momentum.

System \eqref{basic:sde} can be rewritten formally  in the It\^o form:
\begin{eqnarray}
&&d{\frak q}_{y}(t)={\frak p}_y(t)dt
\label{eq:bas}\\
&&\nonumber\\
&& d{\frak p}_y(t)=\left[-(\alpha*{\frak q}(t))_y-\frac{\eps}{2}(\beta*{\frak p}(t))_y\right]dt,\nonumber\\
&&+\sqrt{\eps}\sum_{k=-1,0,1}(Y_{y+k}{\frak p}_y(t))dw_{y+k}(t),\quad y\in\bbZ.\nonumber
\end{eqnarray}
Here 
 $
\beta_y:=\Delta\beta^{(0)}_y,
$
with 
$$
 \beta^{(0)}_y=\left\{
 \begin{array}{rl}
 -4,&y=0\\
 -1,&y=\pm 1\\
 0, &\mbox{ if otherwise.}
 \end{array}
 \right.
 $$
  The lattice Laplacian for a given $g:\bbZ\to\mathbb C$ is defined as
  $\Delta g_y:=g_{y+1}+g_{y-1}-2g_y$. 


 
\subsection{Formulation of the main results}

To describe the distribution of the energy of the chain over the position and momentum coordinates it is convenient to consider  the Wigner transform of the wave function  corresponding to the chain. Adjusting  the time variable to the macroscopic scale it is defined as
\begin{equation}
\label{011307}
\psi^{(\eps)}(t):=\tilde{\om} * {\frak q}\left(\frac{t}{\eps}\right)+i{\frak p}\left(\frac{t}{\eps}\right).
\end{equation}
Here $\tilde{\om}$ is the inverse Fourier transform, see \eqref{inv-fourier}, of   the  dispersion relation function given by
$
\om(k)=\sqrt{\hat \alpha (k)} ,
$ with $\hat\alpha(k)$ the direct Fourier transform  of the potential, defined on  $\bbT$- the one dimensional torus,  see \eqref{fourier}. 
Suppose that  the  initial condition in \eqref{basic:sde} is random, independent of the realizations of the noise and such that for some $\gamma>0$
\begin{equation}
\label{0001}
\limsup_{\eps\to0+}\sum_{y\in\bbZ}\eps^{1+\gamma}\langle|\psi^{(\eps)}_y(0)|^2\rangle_{\eps}<+\infty.
\end{equation}
Here $ \langle\cdot\rangle_{\eps}$ denotes the average with respect to the probability measure $\mu_\eps$ corresponding to the randomness in the initial data. In fact, since the total energy of the system $\sum_{y\in\bbZ}|\psi^{(\eps)}_y(t)|^2$ is conserved in time, see Section 2 of \cite{BOS}, an analogue of condition  \eqref{0001} holds for any $t>0$.

The (averaged) Wigner transform of the wave function, see \cite{BOS}, is a distribution   defined as follows
\begin{eqnarray}
&&
\label{wigner-gam}
\langle W_{\eps,\gamma}(t),\tilde J\rangle
:=\frac{\eps^{1+\gamma}}{2}\sum_{x,x'\in\bbZ}\int_{\bbT}e^{i2\pi(x'-x)k}\tilde J^*\left(\frac{\eps^{1+\gamma}}{2}(x+x'),k\right)\nonumber
\\
&&
\times\bbE_\eps\left[\left( \psi^{(\eps)}_{x'}(t)\right)^* \psi^{(\eps)}_x(t)\right]
\end{eqnarray}
for any  $\tilde J$ belonging to ${\cal S}$ - the Schwartz class  of    functions on   $\bbR\times \bbT$, see Section \ref{sec2.2}.  Here $\bbE_\eps$ is the average with respect to the product measure $\mu_\eps\otimes\bbP$.
%
%
%
 It has been shown in \cite{BOS}, see Theorem  5, that, under appropriate assumptions on the potential $\alpha(\cdot)$, see conditions a1) and a2) below,
 the respective Wigner transforms   $W_{\eps,0}(t)$  converge in a distribution sense, as $\eps\to0+$, to the solution of the linear kinetic equation
\begin{equation}
\label{lattice-wave}
\partial_t U(t,x,k)+\frac{\om'(k)}{2\pi}\partial_x U(t,x,k)={\cal L} U(t,x,k),
\end{equation}
where  ${\cal L}$ is the scattering operator defined in \eqref{011407}.
%
Our principal result concerning this model deals with the   limit of the Wigner transform in the longer time scales, i.e. when $\gamma>0$. It is a direct consequence of Theorems \ref{main-main} and  \ref{main-main1} formulated below that  contain also the information on the convergence rates. Before its statement let us recall the notion of a solution of a fractional heat equation. Assume that $W_0$ is a function from the Schwartz class on $\bbR$. The solution of the Cauchy problem for the fractional heat equation
\begin{equation}
\label{frac-heat}
\partial_tW(t,x)=-\frac{\hat c}{(2\pi)^b}(-\partial_x^2)^{b/2}W(t,x)
\end{equation}
with $b,\hat c>0$ and $W(0,x)=W_0(x)$ is given by
$$
W(t,x):=\int_{\bbR}e^{i2\pi xp-\hat c|p|^{b}t}\hat W_0\left(p\right)dp,
$$
where
$$
\hat W_0(p):=\int_{\bbR}e^{-i2\pi xp}W_0\left(x\right)dx
$$
is the Fourier transform of $W_0(x)$.
\begin{thm}
\label{sketch}
Suppose that potential $\{\alpha_y,\,y\in\bbZ\}$ 
satisfy assumptions $a 1)-a 2)$ formulated in Section \ref{prelim}. 
Then, the following are true.
\begin{itemize}
\item[1)] Assume that $\hat \alpha(0)=0$ (no pinning case), $\gamma\in(0,2a/3)$  for some $a\in(0,1]$ 
and 
\begin{equation}
\label{0001a}
{\cal K}_{a,\gamma}:=\limsup_{\eps\to0+}\eps^{1+\gamma}\int_{\bbT}\langle|\hat \psi^{(\eps)}_0(k)|^2\rangle_{\eps}\frac{dk}{|k|^{2a}}<+\infty,
\end{equation}
where $\hat \psi^{(\eps)}_0(k)$ is the Fourier transform of the initial condition  $ \psi^{(\eps)}_x(0)$.
Suppose also that for some    $W_0$ with the  norm \eqref{norm-ba}  finite  we have
\begin{equation}
\label{011207}
\lim_{\eps\to0+}
\langle W_{\eps,\gamma}(0),\tilde J\rangle
=\int_{\bbR\times\bbT} W_0(x,k)\tilde J^*(x,k)dxdk,
\end{equation}
for  all $\tilde J\in {\cal S}$. Then, for any $t>0$ we have
\begin{equation}
\label{021207}
\lim_{\eps\to0+}
\left\langle W_{\eps,\ga}\left(\frac{t}{\eps^{3\ga/2}}\right),\tilde J\right\rangle
=\int_{\bbR\times\bbT} W(t,x)\tilde J^*(x,k)dxdk,
\end{equation}
where $W(t,x)$ satisfies \eqref{frac-heat} with $b=3/2$ and the initial condition given by
\begin{equation}
\label{init-cond}
W(0,x):=\int_{\bbT}W_0(x,k)dk.
\end{equation}
The coefficient
 $\hat c$ is given by \eqref{031207}.

 \item[2)] Suppose that $\hat \alpha(0)>0$ (pinned case),  $\gamma\in(0,1/2)$ and
conditions \eqref{0001a} and   \eqref{011207}  for    $W_0$ with the  norm \eqref{norm-ba}  finite are satisfied.
  Then, for any  $\tilde J\in{\cal S}$ and $t>0$ we have
\begin{equation}
\label{021207a}
\lim_{\eps\to0+}\left\langle W_{\eps,\ga}\left(\frac{t}{\eps^{2\ga}}\right),\tilde J\right\rangle=\int_{\bbR\times\bbT} W(t,x)\tilde J^*(x,k)dxdk,,
\end{equation}
where  $W(t,x)$ is the solution of the ordinary heat equation, i.e. \eqref{frac-heat} with $b=2$, with the initial condition given by \eqref{init-cond} and coefficient $\hat c$ as in \eqref{031207a}.
\end{itemize}
\end{thm}

\subsection{Fourier transform of the wave function}

\label{prelim}

The one dimensional torus $\bbT$  is understood here as the interval $[-1/2,1/2]$ with its endpoints identified. Let
$e_r(k):=\exp\{-i2\pi rk\}$, $r\in\bbZ$. It is  a standard orthonormal base in $L^2_{\mathbb C}(\bbT)$ - the space of complex valued, square integrable functions. 
The Fourier transform of  a given square integrable sequence of complex numbers $\{g_y,\,y\in\bbZ\}$ is defined as
  \begin{equation}
  \label{fourier}
  \hat g(k)=\sum_{y\in\bbZ}g_ye_y(k), \quad k\in\bbT
\end{equation}
and the  inverse transform is given by
\begin{equation}
\label{inv-fourier}
\tilde f_y=\int_{\bbT} e_y^*(k)f(k)dk
, \quad y\in \bbZ
\end{equation} for any $f$ belonging to $L^2_{\mathbb C}(\bbT)$.

%

A straightforward calculation shows that  $\hat\psi^{(\eps)}(t,k)$ - the Fourier transform  of the  (complex valued)  wave function
given by \eqref{011307} - satisfies the following It\^o stochastic differential   equation, cf. formula (7.0.7) of \cite{BO1},
  \begin{eqnarray}
 \label{basic:sde:2}
&&
 d\hat\psi^{(\eps)}(t)=A_\eps[\hat\psi^{(\eps)}(t)]dt +\sum_{r\in\bbZ}Q[\hat\psi^{(\eps)}(t)](e_r)dw_r(t),\\
 &&
\hat\psi^{(\eps)}(0)= \hat\psi_0,\nonumber
 \end{eqnarray}
where   $ \hat\psi_0$
is the Fourier transform of  $\psi^{(\eps)}(0)$.
Here, 
 a (nonlinear) mapping $A:L^2_{\mathbb C}(\bbT)\to L^2_{\mathbb C}(\bbT)$ is given by
 \begin{equation}
\label{040607}
 A_\eps[f](k):=-\frac{i}{\eps} \om(k)f(k)-\frac{\hat\beta(k)}{4}\sum_{\si=\pm1}\si f_{\si}(k),\quad\forall\,f\in  L^2_{\mathbb C}(\bbT),
 \end{equation}
 where 
 \begin{eqnarray}
   \label{020906}
&&  f_+(k):=f(k)\quad\mbox{and}\quad f_-(k):=f^*(-k),\nonumber
\\
&&
\hat \beta(k)=8\sin^2(\pi k)\left[1+2\cos^2(\pi k)\right].
\end{eqnarray}
For any $g\in L^2_{\mathbb C}(\bbT)$  a linear mapping
 $Q[g]:L^2_{\mathbb C}(\bbT)\to L^2_{\mathbb C}(\bbT)$ is given  by
$$
 Q[g](f)(k):=i\sum_{\si=\pm1}\si\int_{\bbT}r(k,k')g_{\si}(k-k')f(k')dk',\quad\forall\,f\in  L^2_{\mathbb C}(\bbT),
$$
where
 \begin{eqnarray*}
 &&r(k,k'):=\sin(2\pi k)+\sin[2\pi(k-k')]+\sin[2\pi(k'-2k)]
 \\
 &&=4\sin(\pi k)\sin[\pi (k-k')]\sin\left[(2k-k')\pi\right]
 ,\quad k,k'\in \bbT.
 \end{eqnarray*}
Finally, $\{w_r(t),\,t\ge0\}$, $r\in\bbZ$ are  i.i.d. one dimensional, standard Brownian motions, on  a probability space $(\Om,{\cal F},\bbP)$,  that are  non-anticipative w.r.t. the given filtration
$\{{\cal F}_t,\,t\ge0\}$.

We assume, as in  \cite{BOS}, that
 \begin{itemize}
 \item[a1)] $\{\alpha_y,\,y\in\bbZ\}$ is real valued and there exists $C>0$ such that $|\alpha_y|\le Ce^{-|y|/C}$ for all $y\in \bbZ$,
  \item[a2)] $\hat\alpha(k)$ is also real valued,  $\hat\alpha(k)>0$ for $k\not=0$ and in case $\hat \alpha(0)=0$ we  have  $\hat\alpha''(0)>0$.
 \end{itemize}
   The above assumptions imply that both $y\mapsto\alpha_y$ and $k\mapsto\hat\alpha(k)$ are real valued, even functions. In addition $\hat\alpha\in C^{\infty}(\bbT)$ and if $\hat\alpha(0)=0$ then  $\hat\alpha(k)=k^2\phi(k^2)$ for some strictly positive  $\phi\in C^{\infty}(\bbT)$.
This in particular implies that, in the latter case, 
the dispersion relation $\om(k)=\sqrt{\hat \alpha (k)}$ belongs to $C^\infty(\bbT\setminus\{0\})$.

It can be easily checked that under the hypotheses made about the potential $\alpha_y$ the mapping given by \eqref{040607}  is Lipschitz from $L^2_{\mathbb C}(\bbT)$ to itself and  $\sum_{r\in\bbZ}\|Q[g](e_r)\|_{L^2}^2\le C\|g\|_{L^2}^2$ for some $C>0$ and  all $g\in L^2_{\mathbb C}(\bbT)$ so $Q[g]$ is Hilbert-Schmidt. 
 Using   Theorem 7.4, p. 186, of \cite{DZ}  one can show that for any $ L^2_{\mathbb C}(\bbT)$-valued, ${\cal F}_0$-measurable, initial data $\hat\psi_0^{(\eps)}$ there exists a unique solution to \eqref{basic:sde:2} understood as $L^2_{\mathbb C}(\bbT)$ - valued, continuous trajectory, adapted process $\{\hat\psi^{(\eps)}(t),\,t\ge0\}$  a.s. satisfying \eqref{basic:sde:2}. 
In addition, see Section 2 of \cite{BOS},  for every initial data $\hat\psi_0^{(\eps)}\in L^2_{\mathbb C}(\bbT)$ we have
\begin{equation}
\label{conservation}
\|\hat\psi^{(\eps)}(t)\|_{L^2}={\rm const},\quad\forall\,t\ge0,\quad \bbP\mbox{ a.s.} 
\end{equation}

 \subsection{Wigner transform}
 
 \label{sec2.2}
 
 \subsubsection*{Some function spaces}
Denote by  ${\cal S}$ the set of  functions $J:\bbR\times \bbT\to\mathbb C$ that are of $C^\infty$ class and such that for any integers $l,m,n$ we have $\sup_{p,k}(1+p^2)^{n/2}|\partial_p^l\partial_k^mJ(p,k)|<+\infty$. 
For any $a\in\bbR$ we introduce the norm
\begin{eqnarray}
\label{norm-ta0}
&&
\|J\|_{{\cal A}_{a}'}:=\int_{\bbR}(1+p^2)^{a/2}\sup_{k}| J(p,k)|dp,
\end{eqnarray}
for a given  $J\in{\cal S}$. 
By ${\cal A}'_a$ 
we denote the completion of ${\cal S}$ in the norm $\|\cdot\|_{{\cal A}_{a}'}$. 
Note that 
 ${\cal A}_{a}'$ is dual to ${\cal A}_{a}$ defined as the completion of ${\cal S}$ in the norm
 \begin{equation}
\label{012307}
\|J\|_{{\cal A}_a}:=\sup_p(1+p^2)^{-a/2}\int_{\bbT}|J(p,k)|dk.
\end{equation}
We use a shorthand notation ${\cal A}:={\cal A}_0$ and ${\cal A}':={\cal A}_0'$.
%
%
%
With some abuse of notation by $\langle \cdot,\cdot\rangle$ we denote  the scalar product  in $L^2_{\mathbb C}(\bbT)$ and  the extension 
of
$$
\langle J_1,J_2\rangle:=\int_{\bbR\times\bbT}J_1(p,k)J_2^*(p,k)dpdk
$$
from ${\cal S}\times {\cal S}$ to ${\cal A}\times{\cal A}'$.

%
%

We shall also use the space ${\cal B}_{a,b}$ obtained by completion of ${\cal S}$ in the norm 
\begin{eqnarray}
\label{norm-ba}
&&
\|J\|_{{\cal B}_{a,b}}:=\sup_{p}(1+p^2)^{b/2}\int_{\bbT}\frac{| J(p,k)|}{|k|^{2a}}dk.
\end{eqnarray}
When $b=0$ we shall write ${\cal B}_{a}$ instead of ${\cal B}_{a,0}$.

\subsubsection*{Random and average Wigner transform}

For a  given $\eps>0$ let $\hat \psi^{(\eps)}(t)$ be a solution of  \eqref{basic:sde:2} with a random initial condition  $\hat \psi^{(\eps)}(0)$ distributed according to a probability measure $\mu_\eps$ on $L^2_{\mathbb C}(\bbT)$.
Define 
\begin{equation}
\label{011507}
\widehat W_{\eps}(t,p,k):=\left\langle\left(\hat \psi^{(\eps)}\right)^*\left(t,k-\frac{\eps p}{2}\right)\hat \psi^{(\eps)}\left(t,k+\frac{\eps p}{2}\right)\right\rangle_\eps
\end{equation}
and 
 \begin{equation}
 \label{021507}
\widehat Y_{\eps}(t,p,k):=\left\langle\hat \psi^{(\eps)}\left(t,- k+\frac{\eps p}{2}\right)\hat \psi^{(\eps)}\left(t, k+\frac{\eps p}{2}\right)\right\rangle_\eps,
\end{equation}
where, as we recall, $\langle\cdot\rangle_{\eps}$ is the average with respect to the initial condition.
Using   \eqref{0001} and \eqref{conservation}  we conclude that both $\widehat W_{\eps}(t)$ and $\widehat Y_{\eps}(t)$  belong to  $L^1(\bbP;{\cal A})$ - the space of ${\cal A}$-valued random elements possessing the absolute moment,
 for any $t\ge0$.  
We also introduce the average objects $\overline W_\eps(t,p,k)$ and  $\overline Y_\eps(t,p,k)$ using formulas analogous to \eqref{011507} and \eqref{021507} with $\langle\cdot\rangle_{\eps}$ replaced  by $\bbE_\eps$ corresponding to the average over both the initial data and realization of Brownian motion. 

The (averaged) Wigner transform $W_{\eps,\gamma}(t)$   is defined as
   \begin{eqnarray}
 \label{wigner1}
&&
\langle W_{\eps,\gamma}(t),\tilde J\rangle:=\frac{\eps^{1+\ga}}{2}\int_{\bbR\times\bbT}\overline{W}_{\eps}(t,\eps^{\ga}p,k)J^*(p,k)dpdk
\end{eqnarray}
where $J\in{\cal A}'$ and 
$$
\tilde  J(x,k):=\int_{\bbR}\exp\left\{i2\pi px\right\}J(p,k)dp.
$$
The anti-transform $Y_{\eps,\gamma}(t)$ is defined by an analogous formula, with $\overline{W}_{\eps}$ replaced by  $\overline{Y}_{\eps}$.


 \subsection{Evolution of the Wigner transform}
 
 \label{wigner}
Using It\^o formula for the solution of \eqref{basic:sde:2}, see  Theorem 4.17 of \cite{DZ},  we conclude
that
\begin{eqnarray}
\label{wigner-eqt}
&&d\widehat W_\eps(t,p,k)\\
&&=\left\{\left\langle( A_\eps[\hat\psi^{(\eps)}])^*\left(t,k-\frac{\eps p}{2}\right)\hat \psi^{(\eps)}\left(t,k+\frac{p\eps}{2}\right)\right\rangle_\eps\vphantom{\int_0^1}\right.\nonumber\\
&&+\left\langle(\hat\psi^{(\eps)})^*\left(t,k-\frac{\eps p}{2}\right) A_\eps[\hat \psi^{(\eps)}]\left(t,k+\frac{p\eps}{2}\right)\right\rangle_\eps\nonumber\\
&&\left.+\sum_{j\in\bbZ}\left\langle(Q[\hat\psi^{(\eps)}](e_j))^*\left(t,k-\frac{\eps p}{2}\right)Q[\hat\psi^{(\eps)}](e_j)\left(t,k+\frac{\eps p}{2}\right)\right\rangle_\eps\right\}dt
\nonumber\\
&&
+d{\cal M}^{(\eps)}_t(p,k),
\nonumber
\end{eqnarray}
where the $\{{\cal M}^{(\eps)}_t,\,t\ge0\}$ is an $\{{\cal F}_t,\,t\ge0\}$-adapted local martingale, given by 
\begin{eqnarray}
\label{030207}
&&{\cal M}^{(\eps)}_t(p,k)\\
&&
:=\sum_{j\in\bbZ}\int_0^t\left\langle(Q[\hat\psi^{(\eps)}(s)](e_j)^*\left(k-\frac{\eps p}{2}\right)\hat \psi^{(\eps)}\left(s,k+\frac{p\eps}{2}\right)\right\rangle_\eps dw_j(s)\nonumber\\
&&+\sum_{j\in\bbZ}\int_0^t\left\langle(\hat\psi^{(\eps)})^*\left(s,k-\frac{\eps p}{2}\right)Q[\hat \psi^{(\eps)}(s)](e_j)\left(k+\frac{p\eps}{2}\right)\right\rangle_\eps dw_j(s).\nonumber
\end{eqnarray}
In order to guarantee that the  stochastic integrals defined above are martingales, not merely  local ones, we need to make an additional assumption that $\mu_\eps$ has the $4$-th absolute moment.  
Taking the expectation of both sides of \eqref{wigner-eqt} with respect to the realizations of the Brownian motion we conclude that 
$\overline W_\eps(t,p,k)=\bbE\widehat W_\eps(t,p,k)$ satisfies
\begin{eqnarray}
\label{exp-wigner-eqt}
&&\partial_t\overline W_\eps(t,p,k)=-\left[i\delta_{\eps}\om(p,k)+\frac12\bar\beta_{\eps}(p,k)\right]\overline W_\eps(t,p,k)+{\cal R}_\eps^{(0)}(t,p,k)\nonumber \\
&&\!\!\!\!\!\!\!\!\!\!\!\!\!\!\!-4\int_{\bbT}\rho_\eps(k,k-k',p)\bbE\left[(\hat{\frak p}^{(\eps)})^*\left(t,k'-\frac{\eps p}{2}\right)\hat{\frak p}^{(\eps)}\left(t,k'+\frac{\eps p}{2}\right)\right]dk',
\end{eqnarray}
where  
$$
\hat{\frak p}^{(\eps)}(t,k):=\frac{1}{2i}\left[\hat\psi^{(\eps)}(t,k)-(\hat\psi^{(\eps)})^*(t,-k)\right],
$$
$$
\rho_\eps(k,k',p):=r\left(k-\frac{\eps p}{2},k'\right)r\left(k+\frac{\eps p}{2},k'\right),
$$
$$
\delta_{\eps}\om(p,k):=\frac{1}{\eps}\left[\om\left(k+\frac{\eps p}{2}\right)-\om\left(k-\frac{\eps p}{2}\right)\right],
$$
\begin{equation}
\label{010906}
\bar\beta_{\eps}(k,p):=\frac{1}{2}\left[\hat\beta\left(k+\frac{\eps p}{2}\right)+\hat\beta\left(k-\frac{\eps p}{2}\right)\right],
\end{equation}
and
\begin{eqnarray*}
&&
{\cal R}_\eps^{(0)}(t,p,k):=\frac{1}{4}\left\{\hat\beta\left(k-\frac{\eps p}{2}\right)\overline Y_\eps(t, p,k)+\hat\beta\left(k+\frac{\eps p}{2}\right)\overline Y_\eps^*(t, -p,k)\right\}, 
\end{eqnarray*}
with $\overline Y_\eps(t,p,k)=\bbE\widehat Y_\eps(t,p,k)$.

Formula \eqref{exp-wigner-eqt} remains valid also when only the second absolute moment exists. This can be easily argued  by an approximation of the initial condition by random elements that are deterministically bounded.

 Since the momentum, that is the inverse Fourier transform of  $\hat{\frak p}^{(\eps)}(t,k)$, is  real valued, the expression under the expectation appearing on the right hand side of \eqref{exp-wigner-eqt} is an even function of $k'$, thus the last term appearing on the right hand side of the equation can be replaced by 
\begin{equation}
\label{exp-wigner-eqt-1}
-4\int_{\bbT}R_\eps(p,k,k')\bbE_\eps\left[(\hat{\frak p}^{(\eps)})^*\left(t,k'-\frac{\eps p}{2}\right)\hat{\frak p}^{(\eps)}\left(t,k'+\frac{\eps p}{2}\right)\right]dk',
\end{equation}
where
$$
R_\eps(p,k,k'):=\frac12\sum_{\iota=\pm1}\rho_\eps(k,k+\iota k', p).
$$
Note that
 \begin{eqnarray*}
 &&
\!\!\!\!\!\!\!\!R(k,k'):=R_0(p,k,k')=\frac12\left[r^2(k,k-k')+r^2(k,k+k')
\right]\\
&&
=8\sin^2(\pi k)\sin^2(\pi k')\left\{\sin^2\left[\pi (k+k')\right]+\sin^2\left[\pi (k-k')\right]\right\}.\nonumber
 \end{eqnarray*}
 The following relation holds
\begin{equation}
\label{r-beta}
4\int_{\bbT}R(k,k')dk'=\hat \beta(k),\quad\forall\,k\in\bbT.
\end{equation}

We conclude therefore that $\overline W_\eps(t,p,k)$ satisfies the following
\begin{equation}
\label{exp-wigner-eqt-1a}
\langle\partial_t \overline W_\eps(t),J\rangle=
\langle \overline W_\eps(t),\left(iB +{\cal L}\right)J\rangle 
+\langle  {\cal R}_\eps(t),J\rangle,\quad\forall J\in{\cal S}.
\end{equation}
where
$
B f(p,k):=p\om'(k)f(p,k),
$
for any $f\in{\cal S}$. We let $\om'(0):=0$ in case $\om$ is not differentiable at $0$. In addition, 
\begin{equation}
\label{011407}
{\cal L}:={\cal L}^{(0)},
\end{equation}
where for each $\eps\in[0,1]$  operator ${\cal L}^{(\eps)}$ acts on   ${\cal S}$ according to the formula
\begin{eqnarray}
\label{scattering}
&&
{\cal L}^{(\eps)} f(p,k):=2\int_{\bbT}R_\eps(p,k,k')f(p,k')dk'-\frac12\bar\beta_{\eps}(k)f(p,k)\nonumber
\\
&&
=2\int_{\bbT}R_\eps(p,k,k')[f(p,k')-f(p,k)]dk',\quad f\in {\cal S},
\end{eqnarray}
and extends to a bounded operator on either ${\cal A}$, or ${\cal A}'$.
Finally,
\begin{equation}
\label{021606}
{\cal R}_\eps(t,p,k):={\cal R}_\eps^{(1)}(t,p,k)+{\cal R}_\eps^{(2)}(t,p,k),
\end{equation}
with   
\begin{eqnarray}
\label{010207}
&&
{\cal R}_\eps^{(1)}(t,p,k):=\left[i\left(p\om'(k)-\delta_{\eps}\om(p,k)\right)+({\cal L}^{(\eps)} -{\cal L}) \right]\overline W_\eps(t,p,k),\nonumber
\\
&&
{\cal R}_\eps^{(2)}(t,p,k):=\overline{\cal R}_\eps^{(2)}(t,p,k)+\left[\overline{\cal R}_\eps^{(2)}(t,-p,k)\right]^*,
\end{eqnarray}
where
\begin{equation}
\label{020207}
\overline{\cal R}_\eps^{(2)}(t,p,k):=\frac14\hat\beta\left(k-\frac{\eps p}{2}\right)\overline Y_\eps(t, p,k)-\int_{\bbT}R_\eps(p,k,k')\overline Y_\eps(t,p,k')dk'.\nonumber
\end{equation}



Similar calculations show that
\begin{equation}
\label{anti-wigner-eqt}
\frac{d}{dt}\overline Y_\eps(t,p,k)=-\frac{2i}{\eps} \bar\om_\eps(p,k)\overline Y_\eps(t,p,k)
+\overline{\cal U}_\eps(t,p,k),
\end{equation}
where
\begin{eqnarray*}
&&\overline{\cal U}_\eps(t,p,k):=-\frac{1}{2} \bar\beta_\eps(p,k)\overline Y_\eps(t,p,k)-\sum_{\si_1,\si_2=\pm1}\si_1\si_2\int_{\bbT}\bar\rho_\eps(p,k,k')
\\
&&
\times
\bbE_\eps\left\{ \psi^{(\eps)}_{\si_1}\left(t,k-k'-\frac{\eps p}{2}\right)(\psi^{(\eps)}_{\si_2})^*\left(t,k+k'+\frac{\eps p}{2}\right)\right\} dk'
\end{eqnarray*}
and
\begin{eqnarray*}
&&
\bar\om_{\eps}(p,k):=\frac{1}{2}\left[\om\left(k+\frac{\eps p}{2}\right)+\om\left(k-\frac{\eps p}{2}\right)\right],\\
&&
\bar\rho_\eps(p,k,k'):=r\left(k-\frac{\eps p}{2},k'\right)r\left(k+\frac{\eps p}{2},-k'\right).\nonumber
\end{eqnarray*}

\subsection{Probabilistic interpretation of the kinetic linear equation}
\label{sec3.4}
Denote by $\overline U(t,p,k)$ the Fourier transform of the solution of \eqref{lattice-wave} in the $x$ variable. Let
 $K_t(k)$ be a $\bbT$-valued, Markov jump process,  defined over $(\Om,{\cal F},\bbP)$,   starting at $k$,  with the generator ${\cal L}$. 
Suppose also that   $\overline U_0\in{\cal A}$. Then
 \begin{equation}
\label{101307a}
\partial_t \overline U(t)-iB\overline U(t)={\cal L}\overline U(t),\quad \overline U(0)=\overline U_0
\end{equation}
is understood as a continuous ${\cal A}$-valued function $\overline U(t)$ such that
\begin{equation}
\label{101307}
\langle\overline U(t),J\rangle-\langle\overline U_0,J\rangle=\int_0^t\langle\overline U(s),(iB+{\cal L})J\rangle ds
\end{equation}
for all $J\in{\cal A}'$. It is well known that this solution admits the following probabilistic representation 
\begin{equation}
\label{exp-wigner-eqt-1c}
\overline U(t,p,k):=\bbE\left[\exp\left\{- i p\int_0^t\om'\left(K_s(k)\right)ds\right\}\overline U_0\left(p,K_t(k)\right)\right].
\end{equation}
Here $\bbE$ is the expectation over $\bbP$.
 For  given $J\in{\cal A}'$  we let 
\begin{equation}
\label{exp-wigner-eqt-1ca}
J(t,p,k):=\bbE\left[\exp\left\{ i p\int_0^t\om'\left(K_s(k)\right)ds\right\}J\left(p,K_t(k)\right)\right].
\end{equation}
Therefore  $J(t)\in{\cal A}'$. Using the reversibility of the Lebesgue measure under the dynamics of $K_{t}$, we conclude that the law of 
$\{K_{t-s},\,s\in[0,t]\}$ and that of $\{K_{s},\,s\in[0,t]\}$ coincide. Hence,
\begin{equation}
\label{exp-wigner-eqt-1e}
\langle \overline U(t),J\rangle=\langle \overline U_0,J(t)\rangle.
\end{equation}
Likewise, using the definition of ${\cal R}_\eps(t)$ (see \eqref{021606}), from  \eqref{exp-wigner-eqt-1a} and the Duhamel formula we get
\begin{equation}
\label{exp-wigner-eqt-1b}
\langle \overline W_\eps(t),J\rangle=\langle \overline W_\eps(0),J(t)\rangle
+\int_0^t\langle  {\cal R}_\eps(s),J(t-s)\rangle ds,\quad\forall J\in{\cal S}.
\end{equation}

\section{Convergence of the Wigner transform}

 For a given $a\in \bbR$ define the norm
$$
\|f\|_{H^{-a}}:=\left(\int_{\bbT}\frac{|f(k)|^2}{|k|^{2a}}dk\right)^{1/2}.
$$
Recall also that $\mu_{\eps}$ is the distribution of  the initial data for equation \eqref{basic:sde:2}. We assume that:
 
\begin{itemize}
\item[${\rm A}_a$)] for a given $a>0$  
\begin{equation}
\label{W-norm}
{\cal K}_{a,\ga}:=\limsup_{\eps\to0+}\eps^{1+\ga}\int\|f\|_{H^{-a}}^2\mu_{\eps}(df)<+\infty.
\end{equation}
\end{itemize}
Let ${\cal K}_{\ga}:={\cal K}_{0,\ga}$.

\subsection{No pinning}
\label{sec3.2}
 Since
 $$
 \left\langle W_{\eps,\gamma}\left(\frac{t}{\eps^{3\ga/2}}\right),\tilde J\right\rangle=\frac{\eps^{1+\ga}}{2}\int_{\bbR\times\bbT}\overline W_\eps\left(\frac{t}{\eps^{3\ga/2}},p\eps^{\ga},k\right)J^*(p,k) dpdk,
 $$
part 1) of  Theorem \ref{sketch} is a consequence of the following result.
 \begin{thm}
 \label{main-main}
 Assume that   $t_0>0$, $\hat\alpha(0)=0$ and $a\in(0,1]$ is such that \eqref{W-norm} holds. Then,   for any $\gamma\in(0,2a/3)$, 
 \begin{equation}
\label{011005}
0<\ga'< \ga\min\left[\frac{3}{13},\frac{1}{a+1}\right]
\end{equation}
and $b>1$ one can find $C>0$ such that
 \begin{eqnarray}
 \label{051407}
&&
\left|\int_{\bbR\times\bbT}\left[\frac{\eps^{1+\ga}}{2}\overline W_\eps\left(\frac{t}{\eps^{3\ga/2}},p\eps^{\ga},k\right)-\overline W_0(p)e^{-\hat c|p|^{3/2}t}\right]J^*(p,k) dpdk\right|\nonumber\\
&&
\le  \left|\int_{\bbR\times\bbT}\left[\frac{\eps^{1+\ga}}{2}\overline W_\eps\left(0,p\eps^{\ga},k\right)-  W_0\left(p,k\right)\right]e^{-\hat c|p|^{3/2}t}\bar J^*\left(p\right) dpdk\right|\nonumber\\
&&
+ Ct({\cal K}_{\ga,a}+\| W_0\|_{{\cal B}_a})(\|J\|_{{\cal A}_{5}'}+\|J\|_{{\cal B}_{a,b}})\eps^{\ga'}+C{\cal K}_{\ga}\eps\|J\|_{{\cal A}_1'}\left(\frac{t}{\eps^{3\ga/2}}+1\right)\nonumber\\
&&
+C\eps^{2a}\|J\|_{{\cal A}_{2a+ 1}'}\frac{t}{\eps^{3\ga/2}}\left({\cal K}_{\ga}\frac{t}{\eps^{3\ga/2}}+{\cal K}_{a,\ga}\right)
\end{eqnarray}
for all $\eps\in(0,1]$, $\mu_\eps$, $t\ge t_0$, $J\in{\cal A}_5'\cap {\cal B}_{a,b}$  and $W_0\in{\cal B}_a$.
Here 
 \begin{equation}
   \label{010802}
\overline W_0(p)=\int_{\bbT}  W_0\left(p,k\right)dk\quad\mbox{ and  }\overline J(p)=\int_{\bbT} J\left(p,k\right)dk
 \end{equation}
and
 \begin{equation}
   \label{031207}
\hat c:=\left(\frac{\pi^2\hat\al''(0)}{2}\right)^{3/4}.
 \end{equation}
 \end{thm}
\proof
Denote by $\overline  U_\eps\left(t,p,k\right)$ the solution of  \eqref{101307} with the initial condition $\overline  U_\eps\left(0,p,k\right)=W_0\left(p\eps^{-\ga},k\right)$.
The left hand side of \eqref{051407} can be estimated by
 \begin{eqnarray}
 \label{051407a}
&&
\left|\int_{\bbR\times\bbT}\left[\frac{\eps^{1+\ga}}{2}\overline W_\eps\left(\frac{t}{\eps^{3\ga/2}},p\eps^{\ga},k\right)-
\overline  U_\eps\left(\frac{t}{\eps^{3\ga/2}},p\eps^{\ga},k\right)\right]J^*(p,k) dpdk\right| \nonumber
\\
&&
+
\left|\int_{\bbR\times\bbT}\left[
\overline  U_\eps\left(\frac{t}{\eps^{3\ga/2}},p\eps^{\ga},k\right)-\overline W_0(p)e^{-\hat c|p|^{3/2}t}\right]J^*(p,k) dpdk\right|.
\end{eqnarray}
Denote the terms appearing above by ${\cal J}_1$ and ${\cal J}_2$ respectively. The proof is made of two principal ingredients: the estimates of the convergence rates of the averaged Wigner transform of the wave function given by \eqref{basic:sde:2} to the solution of
the linear equation  \eqref{101307a} (that correspond to the estimates of ${\cal J}_1$) and further estimates for the long time-large scale asymptotics of these solutions (these will allow us to estimate ${\cal J}_2$).
To deal with the first issue we formulate the following.
\begin{thm}
 \label{thm-main1} 
Suppose that  $\{\hat\psi^{(\eps)}(t),\,t\ge0\}$ is the solution of \eqref{basic:sde:2} with coefficients satisfying a1)-a2) with $\hat\alpha(0)=0$ and    $A_a$) for some $a\in(0,1]$.  Assume also that $\bar U(t)$ and $J(t)$ are  given by \eqref{exp-wigner-eqt-1c} and \eqref{exp-wigner-eqt-1ca} respectively. Then,
 there exists a constant $C>0$ such that 
\begin{eqnarray}
\label{011606}
&&
\left|\left\langle\frac{\eps^{1+\ga}}{2}\overline  W_\eps(t)-\overline U(t),J\right\rangle\right|\le\left| \left\langle  \frac{\eps^{1+\ga}}{2}\overline W_\eps(0)-\overline U_0,J(t)\right\rangle\right|\nonumber
\\
&&
+ C\eps{\cal K}_{\ga}(t+1)\|J\|_{{\cal A}_{1}'}+C\eps^{2a}t(t{\cal K}_{\ga}+{\cal K}_{a,\ga})\|J\|_{{\cal A}_{2a+1}'},
\end{eqnarray}
for all $J\in {\cal S}$, $\eps\in(0,1]$ and $t\ge0$. 
\end{thm}
We postpone the proof of this theorem until  Section \ref{sec5}, proceeding instead with  estimates of ${\cal J}_1$. 
Using \eqref{011606} we can write that
\begin{eqnarray}
&&
\label{020802}
{\cal J}_1\le \left|\int_{\bbR\times\bbT}dpdk (\delta W)_\eps\left(p,k\right)\right.\\
&&
\left.\times\bbE\left[ J^*\left(p,K_{t/\eps^{3\ga/2}}(k)\right)\exp\left\{- ip\eps^{\ga}\int_0^{t/\eps^{3\ga/2}}\om'(K_s(k))ds\right\}\right] \right|\nonumber\\
&&
+C{\cal K}_{\ga}\eps\|J_\eps\|_{{\cal A}_1'}\left(\frac{t}{\eps^{3\ga/2}}+1\right)+C\eps^{2a}\|J_\eps\|_{{\cal A}_{2a+1}'}\frac{t}{\eps^{3\ga/2}}\left({\cal K}_{\ga}\frac{t}{\eps^{3\ga/2}}+{\cal K}_{a,\ga}\right),\nonumber
\end{eqnarray}
where $J_\eps(p,k):=\eps^{-\ga}J(p\eps^{-\ga},k)$ and
$$
(\delta W)_\eps\left(p,k\right):=\frac{\eps^{1+\ga}}{2}\overline W_\eps\left(0,p\eps^{\ga},k\right)-\overline  W_0\left(p,k\right).
$$ 
Since $\|J_\eps\|_{{\cal A}_b'}\le \|J\|_{{\cal A}_b'}$   for every $\eps\in(0,1]$ and $b\ge0$, we obtain the last two terms on the right hand side of \eqref{020802} account for  the last two terms on the right hand side of \eqref{051407}.

%
Denote the first term on the right hand side of \eqref{020802} by ${\cal I}$. We can write that
${\cal I}\le{\cal I}_1+{\cal I}_2$, where
\begin{eqnarray*}
&&
{\cal I}_1= \left|\int_{\bbR\times\bbT}(\delta W)_\eps\left(p,k\right)e^{-\hat c|p|^{3/2}t}\bar J^*\left(p\right) dpdk\right|,\\
&&
{\cal I}_2 =\left|\int_{\bbR\times\bbT}(\delta W)_\eps\left(p,k\right) \left\{e^{-\hat c|p|^{3/2}t}\bar J^*\left(p\right)\vphantom{\int_0^1}\right.\right.\\
&&
\left.\left.-\bbE\left[J^*\left(p,K_{t/\eps^{3\ga/2}}(k)\right)\exp\left\{-i p\eps^{\ga}\int_0^{t/\eps^{3\ga/2}}\om'(K_s(k))ds\right\}\right]\right\} dpdk\right|.
\end{eqnarray*}
Here for any function $f:\bbR\times \bbT\to\mathbb C$ we  denote
$$
\overline f(p):=\int_{\bbT} f(p,k)dk.
$$ 
Term ${\cal I}_1$ accounts for the first term on the right hand side of   \eqref{051407}.
Using the reversibility of the Lebesgue measure under the dynamics of $K_{t}$ (see \eqref{exp-wigner-eqt-1e}), we conclude that 
\begin{eqnarray*}
&&
{\cal I}_2=\left|\int_{\bbR\times\bbT}J^*\left(p,k\right)\left\{ e^{-\hat c|p|^{3/2}t}\overline {(\delta W)}_\eps(p)\vphantom{\int_0^1}\right.\right.\\
&&
\left.\left.-\bbE\left[(\delta W)_\eps\left(p,K_{t/\eps^{3\ga/2}}(k)\right)\exp\left\{-i p\eps^{\ga}\int_0^{t/\eps^{3\ga/2}}\om'(K_s(k))ds\right\}\right]\right\} dpdk\right|.
\end{eqnarray*}
To estimate ${\cal I}_2$ (and then further to estimate ${\cal J}_2$)  we need a bound on the  convergence rate of the scaled functionals of the form  \eqref{exp-wigner-eqt-1c}.  Let  
\begin{equation}
\label{bar-W}
\overline W(t,p):=\overline W_0(p)\exp\{-\hat c |p|^{3/2}t\}.
\end{equation}
\begin{thm}
For any $t_0>0$, $a\in(0,1]$, $b>1$ and 
$
\ga'$ as in \eqref{011005}
 there exists a constant $C>0$ such that
\label{prop1}
\begin{eqnarray}
\label{010602}
&&
\left|\int_{\bbR\times\bbT}\left\{\overline{ W}(t,p)\vphantom{\int_0^1}-\bbE\left[W_0\left(p,K_{t/\eps^{3\ga/2}}(k)\right)\exp\left\{ -i p\eps^{\ga}\int_0^{t/\eps^{3\ga/2}}\om'(K_s(k))ds\right\}\right]\right\}\right.\nonumber\\
&&\left.\times
J^*(p,k) \vphantom{\int_0^1}dpdk\right|\le   Ct\|W_0\|_{{\cal B}_a}(\|J\|_{{\cal A}_{5}'}+\|J\|_{{\cal B}_{a,b}})\eps^{\ga'},
\end{eqnarray}
for all $\eps\in(0,1]$, $t\ge t_0$, $W_0\in{\cal B}_a$ and $J\in{\cal A}_5'\cap {\cal B}_{a,b}$. 


\end{thm}
The proof of this result shall be presented in Section \ref{sec6.1}.
Using   the above theorem we can estimate 
\begin{eqnarray*}
&&
{\cal I}_2\le Ct(\|J\|_{{\cal A}_{5}'}+\|J\|_{{\cal B}_{a,b}})\eps^{\ga'}\sup_{p}\int_{\bbT}\left[\frac{\eps^{1+\ga}}{2}|\overline  W_\eps\left(0,p\eps^{\ga},k\right)|+|  W_0\left(p,k\right)|\right]\frac{dk}{|k|^{2a}}\\
&&
\le 
Ct({\cal K}_{\ga,a}+\| W_0\|_{{\cal B}_a})(\|J\|_{{\cal A}_{5}'}+\|J\|_{{\cal B}_{a,b}})\eps^{\ga'}.
\end{eqnarray*}
Invoking  again Proposition  \ref{prop1}, this time to estimate  ${\cal J}_2$, we obtain that
$$
{\cal J}_2\le Ct \| W_0\|_{{\cal B}_a}(\|J\|_{{\cal A}_{5}'}+\|J\|_{{\cal B}_{a,b}})\eps^{\ga'}.
$$
The above estimates account for the second term on the right hand side of \eqref{051407}, thus concluding     the proof of the estimate in \eqref{051407}.
\qed

 \subsection{Pinned case}
 
 Part 2) of Theorem \ref{sketch} is a direct consequence of the following result.
  \begin{thm}
 \label{main-main1}
 Assume that $\hat\alpha(0)>0$  and $t_0>0$. Then,   for any $\gamma\in(0,1/2)$, $a\in(0,1]$ and 
 \begin{equation}
\label{011005a}
0<\ga'<\frac{a\ga}{a+1}
\end{equation} 
one can find $C>0$ such that
 \begin{eqnarray}
 \label{051407aa}
&&
\left|\int_{\bbR\times\bbT}\left[\frac{\eps^{1+\ga}}{2}\overline W_\eps\left(\frac{t}{\eps^{2\ga}},p\eps^{\ga},k\right)-\overline W_0(p)e^{-\hat c p^{2}t}\right]J^*(p,k) dpdk\right|\nonumber\\
&&
\le  \left|\int_{\bbR\times\bbT}\left[\frac{\eps^{1+\ga}}{2}\overline W_\eps\left(0,p\eps^{\ga},k\right)-  W_0\left(p,k\right)\right]e^{-\hat c p^{2}t}\bar J^*\left(p\right) dpdk\right|\nonumber\\
&&
+C{\cal K}_{\ga}\eps^{1-2\ga}t\|J\|_{{\cal A}_{ 1}'}
+ Ct({\cal K}_{\ga,a}+\| W_0\|_{{\cal B}_a})(\|J\|_{{\cal A}_{4}'}+\|J\|_{{\cal B}_{a,b}})\eps^{\ga'}\nonumber\\
&&
\end{eqnarray}
for all $\eps\in(0,1]$, $t\ge t_0$, $J\in{\cal A}_{4}'\cap {\cal B}_{a,b}$ and $W_0\in{\cal B}_a$.
Here $\overline W_0(p)$, $\overline J(p)$ and  $
\hat c
$ are respectively given by \eqref{010802} and 
 \eqref{031207a} below.
 \end{thm}

 \proof
 We proceed in the same fashion as in the proof of Theorem \ref{main-main} so we only outline the main points of the argument. First, we estimate the left hand side of \eqref{051407aa} by an expression corresponding to \eqref{051407a}. 
 The first term is estimated by an analogue of Theorem \ref{thm-main1} that in this case can be formulated as follows.
 \begin{thm}
 \label{thm-main1a} 
Assume that  conditions a1)-a2) and $A_0)$ hold. In addition, we let $\hat\alpha(0)>0$. Then, the average Wigner transform $\overline W_\eps(t)$ satisfies the following:  there exists a constant $C>0$ such that 
\begin{equation}
\label{011606a}
\left|\left\langle\frac{\eps^{1+\ga}}{2}\overline  W_\eps(t)-\overline U(t),J\right\rangle\right|\le
\left|\left\langle\frac{\eps^{1+\ga}}{2}\overline  W_\eps(0)-\overline U_0,J(t)\right\rangle\right|
+ C\eps\|J\|_{{\cal A}_{1}'}{\cal K}_{\ga}t
\end{equation}
for all $\eps\in(0,1]$ and $t\ge0$.
\end{thm}
The proof of this result is presented in Section \ref{sec5.2}.
In the next step  we can estimate the rate of convergence of the functional appearing in the formula for the probabilistic solution of the linear Boltzmann equation towards the solution of the heat equation 
 given in the following theorem.  Let 
\begin{equation}
\label{bar-Wa}
\overline W(t,p):=\overline W_0(p)\exp\{-\hat c p^{2}t\}.
\end{equation}
\begin{thm}
For any $t_0>0$, $a\in(0,1]$, $b>1$ and $\ga'$ as in \eqref{011005a}
 there exists a constant $C>0$ such that
\label{prop1a}
\begin{eqnarray}
\label{010602a}
&&
\left|\int_{\bbR\times\bbT}\left\{\overline{ W}(t,p)\vphantom{\int_0^1}-\bbE\left[W_0\left(p,K_{t/\eps^{2\ga}}(k)\right)\exp\left\{ -i p\eps^{\ga}\int_0^{t/\eps^{2\ga}}\om'(K_s(k))ds\right\}\right]\right\}\right.\nonumber\\
&&\left.\times
J^*(p,k) \vphantom{\int_0^1}dpdk\right|\le   Ct\|W_0\|_{{\cal B}_a}(\|J\|_{{\cal A}_{4}'}+\|J\|_{{\cal B}_{a,b}})\eps^{\ga'},
\end{eqnarray}
for all $\eps\in(0,1]$, $t\ge t_0$, $W\in{\cal B}_a$ and $J\in{\cal A}_4'\cap {\cal B}_{a,b}$. 
\end{thm}
The proof of the above theorem is contained in Section \ref{sec6.3}. The remaining part of the argument follows the argument of Section \ref{sec3.2}.

\section{Proofs of Theorems \ref{thm-main1} and \ref{thm-main1a}}

\subsection{Proof of Theorem \ref{thm-main1} }
\label{sec5}


From \eqref{exp-wigner-eqt-1b} and \eqref{exp-wigner-eqt-1e} we conclude that
\begin{eqnarray*}
&&
\left|\frac{\eps^{1+\ga}}{2}\langle \overline W_\eps(t),J\rangle-\langle \overline  U(t),J\rangle\right|\le\left|\left\langle\frac{\eps^{1+\ga}}{2}   W_\eps(0)-\overline U_0,J(t)\right\rangle\right|\\
&&
+\left|\frac{\eps^{1+\ga}}{2}\int_0^t  \langle {\cal R}_\eps(s),J(t-s)\rangle ds\right|,\quad\forall\,\eps\in(0,1],
\end{eqnarray*}
with  ${\cal R}_\eps(t)$  given by \eqref{021606}. Estimates of the last term on the right hand side above shall be done separately for  each term appearing on the right hand side of \eqref{021606}.

\subsubsection{Terms corresponding to  ${\cal R}_\eps^{(1)}$}
\label{R1}
Denote
$$
{\cal E}^{(\eps)}(t,k):=\frac{\eps^{1+\ga}}{2}\overline W_\eps(t,0,k)\quad\mbox{ and }\quad{\cal G}^{(\eps)}(t,k):=\frac{\eps^{1+\ga}}{2}\overline Y_\eps(t,0,k).
$$
\begin{lm}\label{lm011706}
For a given $a\in(0,1]$ there exists $C>0$   such that
\begin{equation}
\label{011906}
\int_{\bbT}\frac{{\cal E}^{(\eps)}(t,k)dk}{|k|^{2a}}\le Ct{\cal K}_{\ga}+{\cal K}_{a,\ga},
\end{equation}
for all $\eps\in(0,1]$, $t\ge0$.
\end{lm}
\proof Denote the expression on the left hand side of \eqref{011906} by ${\cal E}_a(t)$.
From \eqref{exp-wigner-eqt-1a} and  \eqref{021606} we conclude that
\begin{equation}
\label{021906}
\left|\frac{d{\cal E}_a(t)}{dt}\right|
\le \int_{\bbT}\frac{dk}{|k|^{2a}}\left[|{\cal L}({\cal E}^{(\eps)}(t))(k)|+|{\cal L}({\rm Re}\,{\cal G}^{(\eps)}(t))(k)|\right]
\end{equation}
Since $|k|^{-2a}R(k,k')$ is bounded when $a\in(0,1]$ we can bound the right hand side of \eqref{021906}, with the help of  \eqref{conservation}, by  $C{\cal K}_{\ga}$ and \eqref{011906} follows.\qed
%

Let  $\Delta\om_\eps(p,k):=p\om'(k)-\delta_{\eps}\om(p,k)$. For a given $q\in\bbR$ define by ${\rm mod}(q,1/2)$ the unique $r\in [-1/2,1/2)$ such that $q=\ell+r$, where $\ell\in\bbZ$. Divide the cylinder $\bbR\times \bbT$ into two domains: ${\cal C}$ - described below - and its complement ${\cal C}^c$. The first domain  consists of those $(p,k)$ for which either $|{\rm mod }\,(k\pm\eps p/2,1/2)|\ge 1/4$, or both points ${\rm mod }\,(k\pm\eps p/2,1/2)$ belong to an interval $[-1/2,0]$, or they belong to $[0,1/2]$.
We can write  then
\begin{eqnarray*}
&&
\frac{\eps^{1+\ga}}{2}\langle \Delta\om_\eps\overline W_\eps(s),J(t-s)\rangle=I_1+I_2,
\end{eqnarray*}
where the terms on the right hand side correspond to the integration over the aforementioned domains. One can easily verify that there exists $C>0$ such that $|\Delta\om_\eps(p,k)|\le C\eps |p|$ for $(p,k)\in{\cal C}$. We can estimate therefore 
\begin{equation}
\label{012006}
|I_1|\le C\eps\|J\|_{{\cal A}_{1}'}{\cal K}_{\ga}.
\end{equation}
 On the other hand, when $(p,k)\in{\cal C}^c$,
 with no loss of generality assume that $1/2>k+\eps p/2>0>k-\eps p/2>-1/2$ (the other cases can be handled analogously) $p$ is positive and $k\in (-\eps p/2,\eps p/2)$. Since $|\Delta\om_\eps(p,k)|\le C|p|$, for a given $a\in(0,1]$ we can write
 \begin{eqnarray*}
&&
|\Delta\om_\eps(p,k)|\prod_{\si=\pm1}\left|\sin\left[\pi\left(k+\frac{\sigma\eps p}{2}\right)\right]\right|^a\le C\eps ^{2a}|p|^{2a+1}
\end{eqnarray*}
for some constant $C>0$. Using the above estimate we obtain
\begin{eqnarray}
\label{010507}
&&
|I_2|\le C\eps^{2a} \int_{\bbR\times \bbT}|p|^{2a+1}\bbE|J(p,K(t-s,k))|\\
&&
\times \frac{\eps^{1+\ga}}{2} \left\langle\prod_{\sigma=\pm1}\left|\sin\left[\pi\left(k+\frac{\sigma\eps p}{2}\right)\right]\right|^{-a}\left|\widehat\psi^{(\eps)}\left(s,k+\frac{\sigma\eps p}{2}\right)\right|\right\rangle_{\eps}dpdk.\nonumber
\end{eqnarray}
By virtue of \eqref{011906} we get that
\begin{equation}
\label{022006}
|I_2|\le  C\eps^{2a}(s{\cal K}_{\ga}+{\cal K}_{a,\ga})\|J\|_{{\cal A}_{2a+1}'}.
\end{equation}
Taking the Taylor expansion of  $R_\eps(p,k,k')$, up to terms of order $\eps$, it is also straightforward to conclude that
\begin{equation}
\label{032006}
\left|\frac{\eps^{1+\ga}}{2}\langle({\cal L}^{(\eps)} -{\cal L}) \overline W_\eps(s),J(t-s)\rangle\right|
\le \eps {\cal K}_{\ga}\|J\|_{{\cal A}'}.
\end{equation}
Summarizing, \eqref{012006},  \eqref{022006} and \eqref{032006} together imply 
\begin{eqnarray}
\label{042006}
&&
\left|\frac{\eps^{1+\ga}}{2}\int_0^t  \langle {\cal R}_\eps^{(1)}(s),J(t-s)\rangle ds\right|\\
&&
\le C\eps\|J\|_{{\cal A}_{1}'}{\cal K}_{\ga}t+C\eps^{2a}t(t{\cal K}_{\ga}+{\cal K}_{a,\ga})\|J\|_{{\cal A}_{2a+1}'}\nonumber
\end{eqnarray}

\subsubsection{Terms corresponding to  ${\cal R}_\eps^{(2)}$} 
\label{R2}
Straightforward computations, taking the Taylor expansions of  $\hat\beta(k-\eps p/2)$ and $R_\eps(p,k,k')$ up to $\eps$,  show that
\begin{eqnarray*}
&&
\left|\frac{\eps^{1+\ga}}{2}\int_0^t  \langle {\cal R}_\eps^{(2)}(s),J(t-s)\rangle ds-\frac{\eps^{1+\ga}}{2}\int_0^t  \langle {\cal L}( {\rm Re}\,\overline Y_\eps(s)),J(t-s)\rangle ds\right|\\
&&
\le C\eps\|J\|_{{\cal A}'}{\cal K}_{\ga}t.\nonumber
\end{eqnarray*}
From \eqref{anti-wigner-eqt} we conclude the following estimate.
\begin{lm}
\label{lm012206}
Suppose that $\phi_\eps:\bbR\times\bbT\to \bbR$ is such that 
\begin{eqnarray}
\label{anti-wigner-eqt3}
\Gamma_*:=\sup_{\eps,p,k}\frac{|\phi_\eps(p,k)|}{\bar\om_\eps(p,k)}<+\infty. 
\end{eqnarray}
Then, there exists $C>0$ such that
\begin{eqnarray}
\label{012206}
&&\left|\frac{\eps^{1+\ga}}{2}\int_0^t \langle \overline Y_\eps(s)\phi_\eps,J(t-s)\rangle ds\right|\le C\eps {\cal K}_{\ga}(\|J\|_{{\cal A}'_1}t+\|J\|_{{\cal A}'})
\end{eqnarray}
for all $J\in{\cal S}$, $\eps\in(0,1]$, $t>0$.
\end{lm}
\proof
The left hand side of \eqref{012206} can be rewritten as
\begin{eqnarray}
\label{anti-wigner-eqt1}
\left|\frac{\eps^{1+\ga}}{2}\int_0^t\langle\bar\om_\eps\overline Y_\eps(s)\Gamma_\eps,J(t-s)\rangle ds\right|
\end{eqnarray}
where
$
\Gamma_\eps(p,k):=
\phi_\eps(p,k)\bar\om_\eps^{-1}(p,k).$
Using \eqref{anti-wigner-eqt} we can 
estimate this expression by 
\begin{equation}
\label{anti-wigner-eqt5}
\eps\left|\frac{\eps^{1+\ga}}{2}\int_0^t\left\langle\partial_s\overline Y_\eps(s)\Gamma_\eps,J(t-s)\right\rangle ds\right|+\eps\left|\frac{\eps^{1+\ga}}{2}\int_0^t\langle{\cal U}_\eps(s)\Gamma_\eps,J(t-s)\rangle ds\right|.
\end{equation}
Thanks to  \eqref{conservation} and \eqref{anti-wigner-eqt3}
 the second term  is  bounded by $\eps\Gamma_*\|J\|_{\cal A'}{\cal K}_{\ga}t$.
 On the other hand,  integration by parts allows us to estimate the first one by
\begin{eqnarray}
\label{anti-wigner-eqt2}
&&
\eps\left|\frac{\eps^{1+\ga}}{2}\int_0^t\langle\overline Y_\eps(s)\Gamma_\eps,(iB+{\cal L})J(t-s)\rangle ds\right|\\
&&
+\eps\left|\frac{\eps^{1+\ga}}{2}\langle\overline Y_\eps(s)\Gamma_\eps,J(t-s)\rangle \left.\vphantom{\int_0^1}\right|^{t}_{s=0}\right|
\le C{\cal K}_{\ga}\eps(t\|J\|_{{\cal A}_1'}+\|J\|_{{\cal A}'}),\nonumber
\end{eqnarray}
by virtue of  \eqref{conservation} and \eqref{anti-wigner-eqt3}.\qed

Since 
\begin{equation}
\label{010407}
\sup_{\eps,k,k',p}\frac{R_\eps(p,k,k')}{\bar\om_\eps(p,k)}<+\infty
\end{equation}
from the above lemma we conclude that
\begin{equation}
\label{052206}
\left|\frac{\eps^{1+\ga}}{2}\int_0^t  \langle{\cal L}( {\rm Re}\,\overline Y_\eps(s)),J(t-s)\rangle ds\right|
\le C{\cal K}_{\ga}\eps(t\|J\|_{{\cal A}_1'}+\|J\|_{{\cal A}'}).
\end{equation}
This ends the proof of \eqref{011606}.\qed

\subsection{Proof of Theorem \ref{thm-main1a}}
\label{sec5.2}
We maintain the notation from the  argument made in the previous section.  Estimate \eqref{010507} can be improved since in this case there exists  $C>0$ such that
$|\Delta\om_\eps(p,k)|\le C\eps |p|$ for all $(p,k)$, $\eps\in(0,1]$.
Thus, we can write
\begin{equation}
\label{010507a}
|I_2|\le C{\cal K}_{\ga}\eps \int_{\bbR\times \bbT}|p|\bbE|J(p,K(t-s,k))|dpdk\le C{\cal K}_{\ga}\eps \|J\|_{{\cal A}_1'}
\end{equation}
for all $t\ge0$. With this improvement in mind we conclude that there is a constant $C>0$ such that
 \begin{equation}
\label{042006a}
\left|\frac{\eps^{1+\ga}}{2}\int_0^t  \langle {\cal R}_\eps^{(1)}(s),J(t-s)\rangle ds\right|\le C\eps\|J\|_{{\cal A}_{1}'}{\cal K}_{\ga}t
\end{equation}
for all $t\ge0$, $\eps\in(0,1]$.
Repeating the argument from the proof of Theorem  \ref{thm-main1} for the term corresponding to  $ {\cal R}_\eps^{(2)}$ we conclude \eqref{011606a}.
\qed

\section{Convergence rate for a characteristic function of an additive functional of a Markov process}

\subsection{Markov chains}
\label{sec4.1}

Suppose that  $\{\xi_n,\,n\ge0\}$ is a Markov chain  taking values in a Polish metric space $(E, d)$. 
Assume that $\pi$ -  the law of $\xi_0$ - is an invariant and ergodic probability measure for the chain. 
The transition operator satisfies:
\begin{condition}\label{sg}
(spectral gap condition):
$$a:=\sup\lcu \|Pf\|_{L^2(\pi)}:\|f\|_{L^2(\pi)}=1, f\perp\1\rcu< 1.
$$
\end{condition}
Since $P$ is also a contraction in $L^1(\pi)$ and $L^\infty(\pi)$ we
conclude, via Riesz-Thorin interpolation theorem, that for any
$\beta\in[1,+\infty)$:
\begin{equation}
  \label{eq:9}
 \|Pf\|_{L^\beta(\pi)}\le a^{\kappa(\beta)}\|f\|_{L^\beta(\pi)},
\end{equation}
for all $f\in
 L^\beta_0(\pi)$ - the subspace of $
 L^\beta_0(\pi)$ consisting of functions satisfying $\int fd\pi=0$, with $\kappa(\beta):=1-|2/\beta-1|>0$. Thus, $Q_N:=\sum_{n=0}^NP^n$ satisfies
 \begin{equation}
 \label{012606}
 \|Q_Nf\|_{L^{\beta}(\pi)}\le (1-a^{\kappa(\beta)})^{-1} \|f\|_{L^{\beta}(\pi)},\quad \forall \,f\in
 L^\beta_0(\pi).
 \end{equation}
 Furthermore, we assume the following regularity property of transition of probabilities:
 \begin{condition} (existence of bounded probability densities w.r.t. $\pi$)
\label{kernel-a}
transition probability is of the form $P(w,dv)=p(w,v)\pi(dv)$, where the kernel 
$p(\cdot,\cdot)$ belongs to $ L^{\infty}(\pi\otimes \pi).$
\end{condition}


%

\subsection{Convergence of  additive functionals}

Suppose that $\Psi:E\to\bbR$ satisfies the tail estimate
\begin{equation}
\label{phi}
\pi(|\Psi|>\lambda)  \le \frac{ C}{ \lambda^\alpha}
\end{equation}
for some $C>0$ and all $\la\ge1$ and
\begin{equation}
\label{phi-1}
\int \Psi d\pi=0.
\end{equation}
We wish to describe the  behavior of tail probabilities $\P\ls\labs Z_t^{(N)}\rabs\geq N^\ka\rs$ when $\ka>0$ for the scaled partial sum process 
\begin{equation}
\label{050507}
Z^{(N)}_t:=\frac{1}{N^{1/\al}}\sum_{n=0}^{[Nt]}\Psi(\xi_n),\quad t\ge0.
\end{equation}
To that end we represent $Z_t^{(N)}$ as a sum of an $L^\beta$ integrable martingale  for $\beta\in[1,\al)$ and a boundary term vanishing with $N$.
Let $\chi$ be the unique solution, belonging to $L^\beta_0(\pi)$ for $\beta\in[1,\al)$, of the Poisson equation
\begin{equation}
\label{poisson}
\chi-P\chi=\Psi.
\end{equation}
 In fact using Condition \ref{kernel-a} we conclude that $P\chi\in L^\infty(\pi)$. Therefore the tails of $\chi$ and $\Psi$ under $\pi$ are  identical.  We introduce an $L^{\beta}$ integrable martingale letting: $M_0:=0$,
 \begin{equation}
 \label{032906}
 M_N:=\sum_{n=1}^{N}Z_n, \quad \mbox{where }Z_n:=\chi(\xi_{n})-P\chi(\xi_{n-1}),\quad N\geq 1
 \end{equation}
 and the respective partial sum process $M_t^{(N)}:= N^{-1/\al}M_{[Nt]}$, $t\ge0$.
 
 Using the dual version of Burkholder inequality for $L^\beta$ integrable martingales, when $\beta\in(1,2)$, see Corollary 4. 22, p. 101 of \cite{pisier} (and also \cite{jx}) we conclude that there exists $C>0$ such that
\begin{equation}
\label{012306}
\lc\E\labs M_N\rabs^{\beta}\rc^{1/\beta}\leq CN^{1/\beta},\quad\forall\,N\ge1.
\end{equation} 

\begin{lemma}\label{l2}
Under the assumptions \eqref{phi} and \eqref{phi-1} for any $\kappa>0$ and $\delta\in(0,\alpha \kappa)$ there exist $C>0$ such that 
\begin{equation}
\label{032306}
\P\ls\labs Z_t^{(N)}\rabs\geq N^\ka\rs\leq\frac{C( t+1)}{N^\delta},\quad\forall\,N\ge1,\,t\ge 0.
\end{equation}
\end{lemma}
\proof
Choose $\beta\in[1,\al)$.   We can write 
\begin{equation}
\label{022306}
\sum_{n=0}^{[Nt]}\Psi(\xi_n)=M_{[Nt]}+\chi(\xi_0)-\chi(\xi_{[Nt]}).
\end{equation} 
From \eqref{022306}  and Chebyshev's inequality we can estimate the left hand side of \eqref{032306} by
\begin{eqnarray*}
&&\P\ls\labs M_{[Nt]}\rabs\geq\frac{N^{1/\al+\ka}}{3}\rs+\P\ls\labs \chi(\xi_0)\rabs\geq\frac{N^{1/\al+\ka}}{3}\rs+\\
&&
\P\ls\labs \chi(\xi_{[Nt]})\rabs\geq\frac{N^{1/\al+\ka}}{3}\rs\leq\P\ls\labs M_{[Nt]}\rabs\geq\frac{N^{1/\al+\ka}}{3}\rs+\frac{C}{N^{\beta(1/\al+\kappa)}}
\end{eqnarray*}
for some $C>0$.
On the other hand
$$
\P\ls\labs M_{[Nt]}\rabs\geq\frac{N^{1/\al+\ka}}{3}\rs\leq\frac{1}{N^{\beta(1/\al+\ka)}}\E\labs M_{[Nt]}\rabs^{\beta}\le \frac{Ct}{N^{\beta(1/\al+\ka)-1}}.
$$
The last inequality follows from \eqref{012306}. Choosing $\beta$ sufficiently close to $\al$ we conclude the assertion of the lemma. \qed


Suppose furthermore that  an observable  $\Psi:E\to\bbR$ is such that
\begin{condition}\label{tail}
 $\int\Psi d\pi=0$ and
there exist $\alpha \in (1,2)$, $\alpha_1\in(0,2-\alpha)$ and  nonnegative 
  constants $ c_*^+$,$c_*^-$, $C^*>0$ such that $c_*^++c_*^->0$ and
\begin{equation}
\label{psi}
\left|\pi(\Psi>\lambda)    -\frac{ c_*^+}{ \lambda^\alpha}\right|+\left|\pi(\Psi<- \lambda)    -\frac{ c_*^-}{ \lambda^\alpha}\right|\le \frac{C^*}{\lambda^{\alpha+\alpha_1}}
\end{equation}
for all $\lambda\ge1$.
\end{condition}

Let   $\{Z_t,\,t\ge0\}$ be an $\al$-stable process with 
the Levy exponent
\begin{equation}
\label{type-1}
\psi(p):=\alpha\int_{\bbR}(1+i \lambda p-e^{i\lambda
  p})\frac{c_*(\la)d\lambda}{|\lambda|^{1+\alpha}},
\end{equation}
where
\begin{equation}
\label{c-levy}
c_*(\la):=\left\{
\begin{array}{ll}
c_*^-,&\mbox{ when }\la<0,\\
c_*^+,&\mbox{ when }\la>0.
\end{array}
\right.
\end{equation}

In what follows  we shall prove the following.
\begin{thm}\label{main-1}
Under the assumptions made above, for any $\delta\in(0,\al/(\al+1))$  there exist $C>0$ such that 
\begin{equation}
\label{012506}
\labs\E e^{ip Z^{(N)}_t}-e^{-t\psi(p)}\rabs\le C(1+|p|)^{5}(t+1)\left(\frac{1}{N^{\al_1/\al}}+\frac{1}{N^{\delta}}\right)
\end{equation}
for all $p\in\bbR$, $t\ge0$, $N\ge1$. 
\end{thm} 
\proof In the first step we replace an additive functional of the chain by a martingale partial sum process. Using \eqref{022306}  we conclude that there exists  $C>0$ such that 
$$
\labs\E e^{ip Z^{(N)}_t}-\E e^{ip M^{(N)}_t}\rabs\le\frac{C|p|}{N^{1/\al}},\quad\forall\,t\ge0,\,N\ge1,\,p\in\bbR
$$
 and \eqref{012506} shall follow as soon as we can show that  for any $\delta\in(0,\al/(\al+1))$  there exist $C>0$ such that 
  \begin{equation}
\label{022506}
\labs\E e^{ip M^{(N)}_t}-e^{-t\psi(p)}\rabs\le C(1+|p|)^{5}(t+1)\left(\frac{1}{N^{\al_1/\al}}+\frac{1}{N^{\delta}}\right)
 \end{equation}
for all $p\in\bbR$, $t\ge0$, $N\ge1$. 
The remaining part of the argument is therefore devoted to the proof of \eqref{022506}.
Denote  $Z_{N,n}:=N^{-1/\al}Z_n$ with $Z_n$ defined in \eqref{032906} and $\chi$ the solution of  Poisson equation \eqref{poisson} with the right hand side equal to $\Psi$. Introduce also
\begin{eqnarray}
\label{010607}
&&
h_p(x):=e^{ip x}-1-ip x, 
\\
&&
 \bar h_p^{(N)}:=\bbE h_p\lc Z_{N,1}\rc\nonumber
\end{eqnarray}
 and
$
\psi^{(N)}(p):=-N \bar h_p^{(N)}$. Observe that ${\rm Re}\, \bar h_p^{(N)}\le0$. 
\begin{lm}
\label{lm012506}
There exists a constant $C>0$ such that
\begin{equation}
\label{052506}
|\psi^{(N)}(p)-\psi(p)|\le \frac{C}{N^{\alpha_1/\alpha}}|p|(1+|p|).
\end{equation}
In addition, for any bounded set $\Delta\subset\bbR$ and $\beta\in[1,\al)$ there exists $C>0$ such that
\begin{equation}
\label{052506a}
N\int \sup_{\lambda\in\Delta}\left|h_p\left(\frac{\Psi(v)+\lambda}{N^{1/\al}}\right)\right|^{\beta}\pi(dv)\le C|p|^{\beta}(1+|p|),\quad\forall\,p\in\bbR,\,N\ge1.
\end{equation}
\end{lm}
\proof Denote $\tilde \Psi(w,v):=\Psi(v)+P\chi(v)-P\chi(w)$. With this notation the expression on the left hand side of \eqref{052506} can be rewritten as 
\begin{eqnarray}
\label{022706}
&&
\labs N\int_{E\times E} h_p\lc\frac{\tilde \Psi(w,v)}{N^{1/\al}}\rc p(w,v)\pi(dw)\pi(dv)
-\al\int_{\R}h_p(\la)\frac{c_*(\la)}{|\la|^{\al+1}}d\la\rabs\nonumber\\
&&
\le \labs N\int_{E\times E} \left[h_p\lc\frac{\tilde \Psi(w,v)}{N^{1/\al}}\rc-h_p\lc\frac{\Psi(v)}{N^{1/\al}}\rc\right] p(w,v)\pi(dw)\pi(dv)\rabs\nonumber\\
&&
+\labs N\int_{\bbR} h_p\lc\lambda \rc\left[F_N(d\lambda)-\frac{\al c_*(\la)}{N|\la|^{\al+1}}d\la\right] \rabs.
\end{eqnarray}
Here $F_N(\lambda):=\pi(\Psi\le N^{1/\al}\lambda)$.
The first term on the right hand side can be estimated by
\begin{eqnarray*}
&&
  CN^{1-1/\al}\int_{E} \sup_{\lambda\in\Delta}\left|h_p'\lc\frac{\Psi(v)+\lambda}{N^{1/\al}}\rc\right|\pi(dv),
\end{eqnarray*}
where $\Delta$ is a bounded interval containing all possible values of $P\chi(v)-P\chi(w)$.
This expression can be further estimated by
\begin{eqnarray*}
&&
 Cp^2N^{1-2/\al}\int_{E} (\labs\Psi(v)\rabs+1)\pi(dv).
\end{eqnarray*}
As for the second term on the right hand side of \eqref{022706}, using integration by parts, we conclude that it can be bounded by  
\begin{eqnarray}
\label{032706}
&&
N\sum_{\sigma=\pm}\int_{0}^{+\infty} |h_p'(\la)|\left|\pi(\sigma\Psi>N^{1/\al}\lambda)-\frac{c_*^\sigma}{N\la^{\al}}\right|d\la\\
&&
\le C^*|p|N^{-\al_1/\al} \int_{0}^{+\infty} \frac{|e^{ip\la}-1|}{\la^{\al+\al_1}}d\la\le C|p|(1+|p|)N^{-\al_1/\al}.\nonumber
\end{eqnarray}
The first estimate follows from \eqref{psi}, while the last one follows upon the change of variables $\la':=\la p$ and
 the fact that 
 $1<\al+\alpha_1<2$.

Concerning \eqref{052506a} expression appearing there can be estimated by
\begin{eqnarray}
\label{052506b}
&&
N\int_{\bbR} \left|h_p\left(\lambda'\right)\right|^{\beta}F_N(d\lambda')
+\beta N^{1-1/\al} \int_{\bbR} \sup_{\la\in\Delta}\left|h_p\left(\lambda'+\lambda N^{-1/\al}\right)\right|^{\beta-1}
\nonumber\\
&&
\times \left|h_p'\left(\lambda'+\lambda N^{-1/\al}\right)\right|F_N(d\lambda').
\end{eqnarray}
The first term is estimated by
\begin{eqnarray*}
&&
\beta N \int_{0}^{+\infty} \left|h_p\left(\lambda'\right)\right|^{\beta-1}
 \left|h_p'\left(\lambda'\right)\right|\pi(|\Psi|\ge N^{1/\al}\la')d\lambda'\le C|p|^{\beta}
\end{eqnarray*}
for some $C>0$ and all $p\in\bbR$, $N\ge1$. 
Since $\Delta$ is bounded the second term on the other hand is smaller than
\begin{eqnarray}
\label{052506d}
&&
C N^{1-1/\al} \int_{\bbR} \sup_{\la\in\Delta}\left|h_p\left(\lambda'+\lambda N^{-1/\al}\right)\right|^{\beta-1}
 \left|h_p'\left(\lambda'+\lambda N^{-1/\al}\right)\right|F_N(d\lambda')\nonumber\\
&&
\le C N^{1-1/\al}|p|^{\beta+1}\int_{\bbR} \left(|\la'|+\frac{1}{N^{1/\al}}\right)^{\beta}F_N(d\lambda')
\nonumber\\
&&
\le C N^{1-(1+\beta)/\al}|p|^{\beta+1}
\end{eqnarray}
Hence, \eqref{052506a}  follows.
\qed

In fact, in light of the above lemma to prove the theorem it suffices only to show that for any $\delta\in(0,\al/(\al+1))$ one choose $C>0$ so that
\begin{equation}
\label{012506a}
\labs\E e^{ip M^{(N)}_t}-e^{-t\psi^{(N)}(p)}\rabs\le \frac{C}{N^{\delta}}(1+|p|)^{5}(t+1)
\end{equation}
for all $p\in\bbR$, $t\ge0$, $N\ge1$. 
To shorten the notation  we let 
$M_{j,N}:=M_j/N^{1/\al}$.
Since $\{M_n,\, n\ge 0\}$ is adapted and 
$\bbE [ Z_{N,n+1} \,\vert\, {\cal F}_n]=0$, we can write
\begin{eqnarray*}
&& \bbE\Big [e^{ipM_{j+1,N}} \Big] \;=\;
\bbE\Big [e^{ipM_{j,N}} \, 
\bbE\Big [e^{ip  Z_{N,j+1}} 
\, \big\vert\, {\cal F}_j\Big]\, \Big] \\ 
&& \qquad\qquad
\;=\; \bbE\Big [e^{ipM_{j,N}} \Big\{ 1 \;+\; 
\bbE\Big [h_p\left(Z_{N,j+1}\right)
\big\vert\, {\cal F}_j\Big]\, \Big\}\, \Big] \; .
\end{eqnarray*}
To derive a recursive formula for 
$$
W_{j}:=\exp\{\psi^{(N)}(p)j/N\} \bbE  \exp\{ipM_{j,N}\},
$$
 write
\begin{eqnarray*}
&& W_{j+1}-W_j =\exp\{\psi^{(N)}(p)(j+1)/N\}  
\bbE\Big \{e^{ipM_{j,N}}[h_p(Z_{N,j+1})- \bar h_p^{(N)}]\Big\}\\
&& +\exp\{\psi^{(N)}(p)(j+1)/N\} 
\Big( 1+ \bar h_p^{(N)}- e^{ \bar h_p^{(N)}} \Big)
\bbE e^{ipM_{j,N}} \;.
\end{eqnarray*}
Since $M_0=0$, adding up from $j=0$ up to $[Nt]-1$ and then dividing both sides of obtained equality by $\exp\{\psi^{(N)}(p)[Nt]/N\}$
we obtain that
\begin{eqnarray}
\label{01f38}
&&
 \bbE [\exp\{i
pM_t^{(N)}\}] - \exp\{-\psi^{(N)}(p)[Nt]/N\}\\
&& =
\sum_{j=0}^{[Nt]-1}
\bbE\Big \{e_{N,j}[h_p(Z_{N,j+1})- \bar h_p^{(N)}]\Big\}   +\sum_{j=0}^{[Nt]-1}
\Big( 1+ \bar h_p^{(N)}- e^{ \bar h_p^{(N)}} \Big)\bbE e_{N,j}  ,\nonumber
\end{eqnarray}
where
$$
e_{N,j}:=\exp\{\psi^{(N)}(p)(j+1-[Nt])/N\}e^{ipM_{j,N}}.
$$
 We denote the terms appearing  on the right hand side of \eqref{01f38} by $I$ and $I\!I$  and  examine each of them separately. 
As far as   $I\!I$ is concerned  we
bound  its absolute value by
\begin{eqnarray}
\label{062506}
&& Nt\Big| 1+ \bar h_p^{(N)}- e^{ \bar h_p^{(N)}} \Big|
\end{eqnarray}
and since ${\rm Re}\,\bar h_p^{(N)}\le 0$ we obtain  the following
\begin{equation}
\label{072506}
|I\!I|\le Nt\Big|\bar h_p^{(N)} \Big|^2\le \frac{Ct|p|^{2}(1+|p|)^2}{N}
\end{equation}
for some $C>0$.

Fix $K\geq 1$, to be adjusted later on, and divide the set $\Lambda_N = \{0, \dots, [Nt]-1\}$ in
$\ell = [[Nt]/K]+1$ contiguous subintervals, $\ell$ of size $K$ and the last one of size  $K'\le K$, i.e.
\begin{eqnarray*}
&& \Lambda_N \;=\; \bigcup_{m=1}^\ell {\cal I}_m\; , \quad
{\cal I}_m \cap {\cal I}_n \;=\; \varnothing \quad\hbox{for $m\not = n$}\; , \\
&&\quad {\cal I}_m = \{j_m, \dots, j_m + K -1\} , \quad m=1,\ldots,\ell-1
\end{eqnarray*}
and
$$
{\cal I}_{\ell} = \{j_m, \dots, j_m + K' \}
$$
with $K'\le K$.  Here $[a]$ stands for the integer part of
$a\in\bbR$.  To simplify the notation we shall assume that $K'=K$. This assumption does not influence the asymptotics.

We need to estimate the absolute value of
\begin{equation}
\label{020607}
I =\; 
 \sum_{k=1}^{\ell} \sum_{j\in I_k}  \bbE\Big \{e_{N,j}[h_p(Z_{N,j+1})- \bar h_p^{(N)}]\Big\}=I_1+I_2, 
\end{equation}
where
\begin{eqnarray*}
&& I_1 :=\sum_{k=1}^{\ell} \sum_{j\in I_k}\bbE\left\{\left[e_{N,j}-e_{N,j_k} \right]
\bbE\left[h_p(Z_{N,j+1})- \bar h_p^{(N)}\left|\right.{\cal F}_j\right]\right\},
\\
&& I_2 :=\; 
 \sum_{k=1}^{\ell}  \bbE\Big \{e_{N,j_k}\sum_{j\in I_k} [h_p(Z_{N,j+1})- \bar h_p^{(N)}]\Big\},
\end{eqnarray*}
The conditional expectation in the formula for $I_1$ equals
$$
\int p(\xi_j,v)\left[h_p\left(\frac{\Psi(v)+P\chi(v)-P\chi(\xi_j)}{N^{1/\al}}\right)- \bar h_p^{(N)}\right]\pi(dv).
$$
The supremum of its absolute value can be estimated by
$$
2\|p(\cdot,\cdot)\|_{\infty}\int \sup_{\lambda\in\Delta}\left|h_p\left(\frac{\Psi(v)+\lambda}{N^{1/\al}}\right)\right|\pi(dv)\le \frac{C|p|(1+|p|)}{N},
$$
for some constant $C>0$.
Here $\Delta$ is a bounded set containing $0$ and  all possible values of $P\chi(w)-P\chi(z)$ for $z,w\in E$.  The last inequality follows from Lemma \ref{lm012506}.
On the other hand, since the real part of $\log e_{N,j}$ is non-positive we have
$$
\bbE|e_{N,j}-e_{N,j_k}|\le |p|\left(K\bbE |\bar h_p^{(N)}|+\bbE\left[\sup_{l\in I_k}|M_{l,N}-M_{j_k,N}|\right]\right),\quad\forall\,j\in {\cal I}_k.
$$
According to \eqref{052506a} the first term in parentheses  can be estimated by  $C|p|(1+|p|)K/N$. Choose $\beta\in(1,\al)$.
From  \eqref{012306} and Doob's inequality we can estimate the second term by $CK^{1/\beta}/N^{1/\al}$.
Summarizing we have shown that
\begin{equation}
\label{072506aa}
|I_1|\le C(t+1)\left(1+|p|\right)^5\left(\frac{K^{1/\beta}}{N^{1/\al}}+\frac{K}{N}\right).
\end{equation}
On the other hand,
\begin{eqnarray*}
&& I_2 =
 \sum_{k=1}^{\ell}  \bbE\Big \{e_{N,j_k}\sum_{j\in I_k}\bbE\Big [h_p(Z_{N,j+1})- \bar h_p^{(N)}\Big|{\cal F}_{j_k}\Big]\Big\}\\
 &&
 =
 \sum_{k=1}^{\ell}  \bbE\Big \{e_{N,j_k}\Big[\sum_{j=0}^{K-1}P^jg_N(\xi_{j_k})\Big]\Big\},
\end{eqnarray*}
where
$$
g_N(w):=\int h_p\left(\frac{\chi(v)-P\chi(w)}{N^{1/\al}}\right)p(w,v)\pi(dv)- \bar h_p^{(N)}.
$$
Fix $\beta\in(1,\alpha)$. Since $\int g_Nd\pi=0$ and $\ell=[[Nt]/K]+1$, from   \eqref{012606} we conclude  that
\begin{eqnarray*}
&& |I_2| \le (1-a^{\kappa(\beta)})^{-1}\left(\frac{Nt}{K}+1\right)\|g_N\|_{L^{\beta}(\pi)}\\
&&
\stackrel{\eqref{052506a}}{\le} C(1-a^{\kappa(\beta)})^{-1}(t+1)|p|(1+|p|)^{1/\beta}\frac{N^{1-1/\beta}}{K}.
 \end{eqnarray*}
Hence, we have shown that
\begin{eqnarray*}
&& |I| 
\le C(t+1)\left(1+|p|\right)^5\left(\frac{K^{1/\beta}}{N^{1/\al}}+\frac{K}{N}+\frac{N^{1-1/\beta}}{K}\right).
 \end{eqnarray*}
Choose $K=N^{\delta}$ with $\delta\in(0,1)$.
It is easy to see that  
$$
1-\delta>\frac{1}{\al}-\frac{\delta}{\beta},\quad\forall\,\beta\in(1,\al)
$$
thus the middle term on the right hand side is smaller than the first one. The optimal rate 
is obtained therefore when 
$$
 \frac{1}{\al}-\frac{\delta}{\beta}=\delta+\frac{1}{\beta}-1,
$$
or equivalently 
$$
\delta= \frac{1/\al+2}{1/\beta+1}-1.
$$
From the above we get, that the optimal rate of convergence is obtained when $\beta$ is as close to $\al$ as possible, and for each $\delta_1<\al/(\al+1)$ we can choose then $C>0$ so that
\begin{eqnarray*}
&& |I| 
\le \frac{C(t+1)}{N^{\delta_1}}\left(1+|p|\right)^5.
 \end{eqnarray*}
Taking into account the above and \eqref{072506} we conclude  \eqref{012506a}.\qed

\subsection{Convergence rates in the central limit theorem regime}  

\label{berry-essen}

We maintain assumptions made about the chain in the previous section. This time however we assume instead of Condition \eqref{tail} the following.

\begin{condition}\label{taila}
 $\int\Psi d\pi=0$ and
$\Psi\in L^2(\pi)$.
\end{condition}
We define the martingale approximation, using \eqref{032906} from the previous section. Decomposition
\eqref{022306} remains in force. 
Instead of \eqref{012306}
we have then an inequality
 \begin{equation}
\label{012306a}
\lc\E\labs M_N\rabs^{2}\rc^{1/2}\leq CN^{1/2},\quad\forall\,N\ge1,
\end{equation} 
where $C>0$ is independent of $N\ge1$. 
Let 
$$\si^2:= \E M_1^2\quad \mbox{and}\quad 
\psi(p):=\si^2 p^2/2
.
$$ 
Define  the partial sum process $Z^{(N)}_t$ by \eqref{050507} with $\al$ replaced by $2$.
Our main result concerning the convergence of characteristic functions reads as follows.
\begin{thm}\label{main-5}
There exist $C>0$ such that 
\begin{equation}
\label{012506b}
\labs\E e^{ip Z^{(N)}_t}-e^{-t\psi(p)}\rabs\le C(1+|p|)^4\frac{t+1}{N^{1/3}}
\end{equation}
for all $p\in\bbR$, $t\ge0$, $N\ge1$. 
\end{thm}
\proof
Denote  $Z_{N,n}:=N^{-1/2}Z_n$ with $Z_n$ defined in \eqref{032906} and $M_{j,N}:=M_j/N^{1/2}$. 
Using this notation as well as the notation from the previous section we can write that
$$
\labs\E e^{ip Z^{(N)}_t}-\E e^{ip M^{(N)}_t}\rabs\le\frac{C|p|}{N^{1/2}},\quad\forall\,t\ge0,\,N\ge1,\,p\in\bbR.
 $$
It suffices therefore to prove 
\begin{equation}
\label{012506c}
\labs\E e^{ip M^{(N)}_t}-e^{-t\psi(p)}\rabs\le C(1+|p|)^4\frac{t+1}{N^{1/3}}
\end{equation}

Let
$h_p(x)$ be given by  \eqref{010607}.
We denote  
 $\bar h_p^{(N)}:=-(\si p)^2/(2N)$ and
$$
W_{j}:=\exp\{(\si p)^2j/(2N)\} \bbE [ \exp\{ipM_{j,N}\}],
$$
Analogously to what has been done in the previous section we obtain
\begin{eqnarray*}
&& W_{j+1}-W_j =\exp\{(\si p)^2((j+1)/(2N))\}  
\bbE\Big \{e^{ipM_{j,N}}[h_p(Z_{N,j+1})- \bar h_p^{(N)}]\Big\}\\
&& +\exp\{(\si p)^2((j+1)/(2N))\} 
\Big( 1+ \bar h_p^{(N)}- e^{ \bar h_p^{(N)}} \Big)
\bbE e^{ipM_{j,N}}\;.
\end{eqnarray*}
Adding up from $j=0$ up to $[Nt]-1$ and then dividing both sides of obtained equality by $\exp\{(\si p)^2[Nt]/(2N)\}$
we obtain that
\begin{eqnarray}
\label{01f38a}
&&
 \bbE [\exp\{i
pM_t^{(N)}\}] - \exp\{-(\si p)^2[Nt]/(2N)\}\\
&& =
\sum_{j=0}^{[Nt]}
\bbE\Big \{e_{N,j}[h_p(Z_{N,j+1})- \bar h_p^{(N)}]\Big\}  +\sum_{j=0}^{[Nt]}
\Big( 1+ \bar h_p^{(N)}- e^{ \bar h_p^{(N)}} \Big)\bbE e_{N,j},\nonumber
\end{eqnarray}
where
$$
e_{N,j}:=\exp\{(\si p)^2(j+1-[Nt])/(2N)\}e^{ipM_{j}/N^{1/2} }.
$$
Denote the absolute value of the  terms appearing  on the right hand side by $I$ and $I\!I$ respectively. We can easily estimate 
$$
I\!I\le C\frac{tp^4}{N}
$$
for some constant $C>0$ and all $t\ge0$, $N\ge1$ and $p\in\bbR$.

To estimate $I$ we invoke the  block argument from  the previous section. Fix $K\geq 1$ and divide the set $\Lambda_N = \{0, \dots, N-1\}$ in
$\ell = [[Nt]/K]+1$ contiguous subintervals, $\ell$ of size $K$ and the last one of size  $K'\le K$. In fact for simplicity sake we just assume that all intervals have the same size and maintain the notation introduced in the previous section. Estimate \eqref{020607} remains in force with the obvious adjustments needed for $\alpha=2$. Repeating the argument leading to \eqref{072506aa} with \eqref{012306a} used in place of \eqref{012306} we conclude that
\begin{equation}
\label{072506a}
|I_1|\le C p^2(t+1)\left(\frac{pK^{1/2}}{N^{1/2}}+\frac{p^2K}{N}\right).
\end{equation}
On the other hand,
\begin{eqnarray*}
&& |I_2| \le (1-a)^{-1}\left(\frac{Nt}{K}+1\right)\|g_N\|_{L^{2}(\pi)}
\le C(1-a)^{-1}\frac{(t+1)p^{2}}{K}.
 \end{eqnarray*}
Choosing $K=N^{1/3}$ in the above estimates we conclude \eqref{012506c}.

\subsection{Convergence rates for additive functionals of jump processes}

\label{jump-process}

Assume that $\{\tau_n, \,n\geq 0\}$ are i.i.d. exponentially distributed random variables  with $\bbE\tau_0=1$ that are independent of the Markov chain $\{\xi_n,\,n\ge0\}$ considered in the previous section. Suppose furthermore that $V, \theta$ are Borel measurable functions on $E$ such that:
\begin{condition}\label{lower-t}
For some $t^*>0$
\begin{equation}
\label{022611}
\theta(w)\ge t_*,\mbox{\ \ \ } \forall\ w\in E,
\end{equation}
there exist   $C^*>0$ and $\alpha_2>1$ such that 
\begin{equation}
\label{theta}
\pi(\theta>\lambda) \le \frac{C^*}{\lambda^{\alpha_2}},\quad\forall\, \lambda\ge1.
\end{equation}
\end{condition}
We shall also assume that
either
$
\Psi(w):=V(w)\theta(w)$ satisfies Condition \ref{tail},
or
it satisfies  Condition \ref{taila}.

Let $t_n:=\sum_{k=0}^{n-1}\theta(\xi_k)\tau_k$ and let $\{X_t,\,t\ge0\}$ be a jump process given by
$X_t:=\xi_{n_t}$
where
$
n_t:=n\mbox{ \ for \ }t_n\leq t<t_{n+1}$. The process conditioned on  $\xi_0=w$ shall be denoted by $X_t(w)$. 
The corresponding chain shall be called a skeleton, while $\theta^{-1}(w)$ is the jump rate at  $w$. Note that measure $\tilde \pi(dw):=\bar\theta^{-1}\theta(w)\pi(dw)$ is invariant under the process. Here $\bar \theta:=\E \theta(\xi_0)$.

We also define processes 
\begin{equation}
\label{Y-N}
Y^{(N)}_t:=N^{-1/\beta}\int_0^{Nt}V(X_s)ds,
\end{equation}
with $\beta=\alpha$ when this condition holds and
$\beta=2$, in case Condition \ref{taila} is in place.

Observe that the integral above can be written as a random length sum formed over a Markov chain. More precisely, when \eqref{psi} holds
\begin{equation}
\label{012906}
Y^{(N)}_t=\frac{1}{N^{1/\al}}\sum_{k=0}^{n_{Nt}-1}\tilde \Psi(\tilde\xi_k)+R^{(N)}_t,
\end{equation}
where
\begin{equation}
\label{012906a}
R^{(N)}_t:=\frac{1}{N^{1/\al}}V(\xi_{n_{Nt}})(t-t_{n_{Nt}}),
\end{equation}
$\{\tilde \xi_n,\,n\ge0\}$ is an $E\times(0,+\infty)$-valued Markov chain   given by $\tilde\xi_n:=(\xi_n,\tau_n)$ and $\tilde\Psi(w,\tau):=\Psi(w)\tau$. Its invariant measure $\tilde \pi$ is given by $\tilde\pi(dw,d\tau):=e^{-\tau}\pi(dw)d\tau$ and the transition probability kernel equals
$
\tilde P(w,\tau,dv,d\tau'):=p(w,v)e^{-\tau'}\pi(dv)d\tau'.
$
It is elementary to verify that this chain and the observable $\tilde \Psi$ satisfy Conditions \ref{sg}-\ref{tail} with the constants appearing in Condition \ref{tail} given by $\tilde c_*^{\pm}:=\Gamma(\al+1) c_*^{\pm}$, where $\Gamma(\cdot)$ is the Euler gamma function.
In case Condition \ref{taila} holds we can still write \eqref{012906} and \eqref{012906a} with $\alpha=2$.

 Consider the stable process $\{Z_t,\,t\ge0\}$ whose Levy exponent is given by \eqref{type-1} with $c_*(\la)$ replaced by 
\begin{equation}
\label{c-levy-p}
\tilde c_*(\la):=\left\{
\begin{array}{ll}
\bar\theta^{-\al}\Gamma(\al+1)c_*^-,&\mbox{ when }\la<0,\\
\bar\theta^{-\al}\Gamma(\al+1)c_*^+,&\mbox{ when }\la>0.
\end{array}
\right.
\end{equation}
Let $\{B_t,\,t\ge0\}$  be a zero mean  Brownian motion whose variance equals 
\begin{equation}
\label{100607}
\hat c^2:=2\bar\theta^{-2}\si^2.
\end{equation}
The aim of this section is to prove the following.
\begin{thm}
\label{main}
In addition to the assumptions made in Section \ref{sec4.1} suppose that Condition \ref{lower-t}  holds. Then, for any
 \begin{equation}
 \label{delta*}
 0<\delta<\delta_*:=\min[\al/(\al+1),(\al_2-1)/(\al\al_2+1)]
 \end{equation}
  there exists $C>0$ such that 
\begin{equation}
\label{012611}
\left|\E\exp\left\{i p Y^{(N)}_t\right\}-\E\exp\left\{ip Z_t\right\}\right|\le C(|p|+1)^5(t+1)\left(\frac{1}{N^{\al_1/\al}}+\frac{1}{N^{\delta}}\right),
\end{equation}
for all $p\in\bbR$, $t\ge 0$, $N\ge1$.

If, on the other hand  the assumptions made in Section \ref{berry-essen}   and Condition \ref{lower-t}  hold then  for any 
\begin{equation}
\label{delta*1}
0<\delta<\delta_*:=(\al_2-1)/(1+2\al_2)
\end{equation}
 there exists $C>0$ such that 
\begin{equation}
\label{012611a}
\left|\E\exp\left\{i p Y^{(N)}_t\right\}-\E\exp\left\{ip B_t\right\}\right|\le C(|p|+1)^4(t+1)\left(\frac{1}{N^{1/3}}+\frac{1}{N^{\delta}}\right)
\end{equation}
for all $p\in\bbR$, $t\ge 0$, $N\ge1$.
\end{thm}
The proof of this result is carried out below. It relies on  Theorem \ref{main-1}. 
The principal difficulty is that definition of $Y^{(N)}_t$ involves a random sum instead of a deterministic one as in the previous section. This shall be resolved by replacing $n_{Nt}$ appearing in the upper limit of the sum  in \eqref{012906} by the deterministic limit $[\bar Nt]$, where  $\bar N:=N/\bar \theta$, that can be done with a large probability, according to the lemma formulated below. 
\subsubsection{From a random to deterministic sum}
\begin{lemma}\label{l1}
Suppose that $\kappa>0$. Then, for any $\delta\in(0,\al_2 \ka)$ there exists $C>0$ such that
\begin{equation}
\label{eq3}
\P\ls\labs n_{ N t}-[\bar N t]\rabs\geq 
N^{\ka+1/\al_2}\rs\leq\frac{C(t+1)}{N^\delta},\quad\forall\,N\ge1,\,t\ge 0.
\end{equation}
\end{lemma}
\proof
Denote 
$$\quad $$
\begin{eqnarray}
\label{052906}
&&
A_N^+:=\ls n_{[ N  t]}-[\bar N t]\geq N^{1/\al_2+\ka}\rs,\\
&&
A_N^-:=\ls n_{[ N t]}-[\bar N t]\leq -N^{1/\al_2+\ka}\rs,\quad
A_N:=A_N^+\cup A_N^-.\nonumber
\end{eqnarray}
Let  $\kappa'\in(0,\kappa)$ be arbitrary and 
$$C_N:=\ls \sum_{n=[\bar N t]}^{[\bar N t]+N^{1/\al_2+\ka}-1}\tau_n\geq N^{1/\al_2+\ka'}\rs.$$
We adopt the convention of summing up to the largest integer smaller than, or equal to the upper limit of  summation. 
Note that  on $A_N^+$ we have
$$
[ N t]-t_{[\bar N t]}\geq t_{n_{[N t]}}-t_{[\bar N t]}\geq t^*\sum_{n=[\bar N t]}^{[\bar N t]+N^{1/\al_2+\ka}-1}\tau_n.$$
Furthermore
$$
r_N:=\frac{1}{N^{1/\al_2}}\left|\frac{[Nt]}{\bar\theta}-[\bar N t]\right|\le\frac{1}{N^{1/\al_2}}\left(\frac{1}{\bar\theta}+1\right).
$$
Hence,
\begin{eqnarray*}
&&
\P\ls A_N^+\cap C_N\rs\leq\P\ls\labs[ N t]-t_{[\bar N t]}\rabs\geq t^*N^{1/\al_2+\ka'}\rs\\
&&
\le \P\ls\frac{1}{N^{1/\al_2}}\labs\sum_{n=0}^{[\bar N t]}[ \theta(\xi_n)\tau_n-\bar \theta]\rabs\geq t^*N^{\ka'}-\bar\theta r_N\rs.
\end{eqnarray*}
To estimate the  probability appearing on the utmost right hand side we apply Lemma \ref{l2} for the Markov chain $\{\tilde \xi_n,\,n\ge0\}$ and $\Psi(w,\tau):=\theta(w)\tau-\bar \theta$.
We conclude therefore that for any $\delta\in(0,\al_2 \ka')$ there exists $C>0$ such that
\begin{equation}
\label{042906}
\P\ls A_N^+\cap C_N\rs\leq \frac{C(t+1)}{N^{\delta}},\quad\forall\,t\ge0,\,N\ge1.
\end{equation}
Since $\bbE\tau_0=1$ and $\ka'\in(0,\ka)$   for any $x\in(0,1)$  we can find $C>0$ such that
\begin{eqnarray}\label{eq2}
&&
\P\ls C_N^c\rs\leq\P\ls\frac{1}{N^{1/\al_2+\ka}}\sum_{n=0}^{N^{1/\al_2+\ka}-1}\tau_n<\frac{1}{N^{\ka-\ka'}}\rs\\
&&
\le C \P\ls\frac{1}{N^{1/\al_2+\ka}}\sum_{n=0}^{N^{1/\al_2+\ka}-1}\tau_n<x\rs\le C \exp\lcu-N^{1/\al_2+\ka}I(x)\rcu,\quad\forall\,N\ge 1\nonumber
\end{eqnarray}
 where $I(x):=-(1-x-\ln x)$. The last inequality follows from the large deviations estimate of Cramer, see e.g. Theorem 2.2.3 of \cite{dembo-zeitouni}.
Using this and \eqref{042906} we get
$$\P\ls A_N^+\rs\leq\P\ls A_N^+\cap C_N\rs+\P\ls C_N^c\rs\leq\frac{C}{N^\delta}.$$
Probability  $\P[A_N^-]$ can be estimated in similar way.  Instead of $C_N$ we consider the event
$$
\tilde C_N:=\ls \sum_{n=[\bar N t]-N^{1/\al_2+\ka}+1}^{[\bar N t]}\tau_n\geq N^{1/\al_2+\ka'}\rs
$$
 and carry out similar estimates to the ones done before.
\qed

\subsubsection{Proof  of \eqref{012611}}
Choose any $\kappa>0$.
We can write 
\begin{eqnarray}
\label{060607}
&&
\left|\E\exp\left\{i p Y^{(N)}_t\right\}-\E\exp\left\{ip Z_t\right\}\right|\\
&&
\le 
\left|\E\left[\exp\left\{i p Y^{(N)}_t\right\}-\exp\left\{ip Z_t^{(\bar N)}\right\},A_N\right]\right|\nonumber\\
&&
+\left|\E\left[\exp\left\{i p Y^{(N)}_t\right\}-\exp\left\{ip Z_t^{(\bar N)}\right\},A_N^c\right]\right|\nonumber\\
&&+\left|\E\exp\left\{i p Z^{(\bar N)}_t\right\}-\E\exp\left\{ip Z_t\right\}\right|,\nonumber
\end{eqnarray}
with $A_N$ defined in \eqref{052906}. The last term on the right hand side can be estimated by the expression
appearing on the right hand side of \eqref{012611}, by virtue of Theorem \ref{main-1}.

The first term on the right hand side can be estimated by
$
C(t+1)N^{-\delta}
$
for some $\delta\in(0,\al_2 \ka)$ and  $C>0$. The second term is less than, or equal to
$$
 |p|(\E I_N+\E J_N),
$$
where
\begin{equation}
\label{070607}
I_N:=N^{-1/\al}\max_{m\in[[\bar N t]-N^{1/\al_2+\ka},[\bar N t]+N^{1/\al_2+\ka}]\cap\Z}\labs\sum_{k=m}^{[\bar N t]}\tilde\Psi(\tilde \xi_k)\rabs,
\end{equation}
and
\begin{equation}
\label{080607}
J_N:=N^{-1/\al}\max_{m\in[[\bar N t]-N^{1/\al_2+\ka},[\bar N t]+N^{1/\al_2+\ka}]\cap\Z}\labs\tilde \Psi(\tilde \xi_m)\rabs.
\end{equation}
\begin{lm}
\label{lm032611}
Suppose that $\kappa\in(0,1-1/\al_2)$. Then, for any $\delta\in(0,\al^{-1}(1- \ka-\al^{-1}_2))$ there exists $C>0$ such that
\begin{equation}
\label{122611}
\bbE I_N\le \frac{C}{N^{\delta}}
\end{equation}
and
\begin{equation}
\label{132611b}
\bbE J_N\le \frac{C}{N^{\delta}},\quad\forall\,N\ge1.
\end{equation}
\end{lm}
\proof
First we prove \eqref{132611b}. Since $\tilde \Psi(\tilde\xi_0)$ is $L^{\beta}$ integrable 
 we can write for any $\beta\in(1,\al)$
\begin{eqnarray}
\label{062906}
&&\bbE J_N\le 
\frac{1}{N^{1/\al}} \lcu\E\max_{m\in [[\bar N t]-N^{1/\al_2+\ka},[\bar
  N t]+N^{1/\al_2+\ka}]\cap\N}|\tilde \Psi(\tilde\xi_m)|^{\beta}\rcu^{1/\beta}\\
&&
\le\frac{1}{N^{1/\al}} \lcu\E\sum_{m\in  [[\bar N t]-N^{1/\al_2+\ka},[\bar
  N t]+N^{1/\al_2+\ka}]\cap\N}|\tilde \Psi(\tilde\xi_m)|^{\beta}\rcu^{1/\beta}
\le CN^{(1/\al_2+\ka)/\beta-1/\al}. \nonumber
\end{eqnarray}
Choosing $\beta$ sufficiently close to $\al$ we conclude \eqref{132611b}.

Now we prove \eqref{122611}.
Again we can use martingale decomposition
$$
\sum_{n=0}^{m-1}\tilde \Psi(\tilde \xi_k)=\tilde \chi(\tilde \xi_0)-\tilde P\tilde \chi(\tilde \xi_{m-1})+M_m,
$$
where 
$$
M_m:=\sum_{n=1}^{m-1}\left[\tilde \chi(\tilde \xi_{n})-\tilde P\tilde \chi(\tilde \xi_{n-1})\right],
$$
 and $\tilde \chi(\cdot)$ is unique, zero mean, solution of $\tilde \chi-\tilde P\chi=\tilde \Psi$, with $\tilde P$ the transition operator for the chain $\{\tilde \xi_n,\,n\ge0\}$. 
Using stationarity  we can bound
\begin{eqnarray*}
&&\E I_N\leq\frac{2}{N^{1/\al}}\E \max_{m\in\ls 0,2N^{1/\al_2+\ka}\rs\cap\Z}\labs\sum_{n=0}^{m-1}\tilde \Psi(\tilde \xi_k)\rabs\leq\frac{2}{N^{1/\al}}\E\max_{m\in[0,2N^{1/\al_2+\ka}]\cap\Z}\labs M_m\rabs\\
&&
+
\frac{2}{N^{1/\al}}\E \labs\tilde\chi(\tilde\xi_0)\rabs+
\frac{2}{N^{1/\al}}\E\max_{m\in[0,2N^{1/\al_2+\ka}]\cap\Z} \labs\tilde\chi(\tilde\xi_m)\rabs.
\end{eqnarray*}
Denote the terms on the right hand side by $I_N^{(i)}$, $i=1,2,3$
respectively.
One can easily estimate
$
 I_N^{(2)}\le CN^{-1/\al}.
$
Also, to bound $I_N^{(3)}$ we can repeat estimates made in 
\eqref{062906}, as $\tilde \chi$ is also $ L^{\beta}$ integrable and obtain
$$
 I_N^{(3)}\le CN^{(1/\al_2+\ka)/\beta-1/\al}
$$
for some $C>0$. Finally, to deal with $I_N^{(1)}$ observe that
by Doob's inequality for $\beta\in(1,\al)$
\begin{eqnarray*}
&&
\frac{1}{N^{1/\al}} \bbE\max_{m\in [0,2N^{1/\al_2+\ka}]\cap\Z}\left|M_m\right|\le\frac{C}{N^{1/\al} }\left\{\bbE\left|M_{2N^{1/\al_2+\ka}}\right|^\beta\right\}^{1/\beta}.
\end{eqnarray*}
We use again \eqref{012306} and conclude that
$$
\left\{\bbE\left|M_{2N^{1/\al_2+\ka}}\right|^\beta\right\}^{1/\beta}\le CN^{(1/\al_2+\ka)/\beta}.
$$
Summarizing from the above estimates we get
$$
I_N^{(1)}\le CN^{-\delta},
$$
where $\delta$ is as in the statement of the lemma.
\qed

Gathering the above results we have shown that for any $\kappa\in(0,1-\al^{-1}_2)$ and $\delta_1\in (0,\al_2 \kappa)$, $\delta_2\in(0,\al^{-1}(1- \ka-\al^{-1}_2))$ we have
$$
\left|\E\exp\left\{i p Y^{(N)}_t\right\}-\E\exp\left\{ip Z_t^{(\bar N)}\right\}\right|\le C\left(\frac{t+1}{N^{\delta_1}}+\frac{|p|}{N^{\delta_2}}\right).
$$
Choosing $\ka$ sufficiently close to $(1-\al^{-1}_2)/(\al\al_2+1)$ we obtain that for any  $\delta\in(0,(\al_2-1)(\al\al_2+1)^{-1})$ we can find a constant $C>0$ so that
\begin{equation}
\label{072906}
\left|\E\exp\left\{i p Y^{(N)}_t\right\}-\E\exp\left\{ip Z_t^{(\bar N)}\right\}\right|\le \frac{C}{N^{\delta}}(t+1)(|p|+1).
\end{equation}
Thus, we conclude the proof of \eqref{012611}.

 \subsubsection{Proof  of \eqref{012611a}}
 
 In this case we can still write  inequality  \eqref{060607}. With the help of Lemma \ref{l1}, for any $\kappa>0$ and $\delta\in(0,\al_2 \ka)$ we can find $C>0$ such that  
 \begin{equation}
\label{060607a}
\left|\E\exp\left\{i p Y^{(N)}_t\right\}-\E\exp\left\{ip Z_t^{(\bar N)}\right\}\right|
\le \frac{C(t+1)}{N^{\delta}}+
 |p|(\E I_N+\E J_N),
\end{equation}
 where $I_N$, $J_N$ are defined by \eqref{070607} and \eqref{080607} respectively, with $\al=2$.
 repeating the estimates made in the previous section we obtain that
 $$
 \E I_N+\E J_N\le CN^{1/2(\ka-1+\al^{-1}_2)}
 $$
 for some $C>0$. Using the above estimates and  \eqref{012506b} we conclude   \eqref{012611a}.

\section{Proofs of Theorems \ref{prop1} and \ref{prop1a}}

\subsection{Proof of Theorem \ref{prop1}}

\label{sec6.1}

 Let $N:=\eps^{-3\ga/2}$ 
and $J\in {\cal A}$ be a real valued function. Define
\begin{equation}
\label{bar-U}
  W_N(t,p,k):=\E\left[W\lc p,K_{Nt}(k)\rc\exp\left\{ -i pN^{-2/3} \int_0^{Nt}\om'(K_s(k))ds, \right\}\right].
\end{equation}
where $\{K_t(k),\,t\ge0\}$ is the Markov jump process starting at $k$, introduced in Section \ref{sec3.4}. It
 can be easily verified that the Lebesgue measure on the torus is invariant and reversible for the process and we denote
 the respective stationary process by  $\{K_t,\,t\ge0\}$.
Its generator 
  ${\cal L}$ is a symmetric operator on $L^2(\bbT)$ given by
 \begin{eqnarray*}
&& {\cal L}f(k):=\int_{\bbT} R(k,k')[f(k')-f(k)]dk' \\
&&
 =-R(k)f(k)+\frac{3}{4}\sum_{\iota\in\{-1,1\}}\langle \frak e_\iota,f\rangle \frak e_{-\iota}(k),\quad \forall\, f\in L^2(\bbT).
\end{eqnarray*}
Here 
$$
\frak{e}_1(k):=\frac{8}{3}\sin^4(\pi k),\quad \frak{e}_{-1}(k):=2\sin^2(2\pi k).
$$
We also let 
$$
\frak r(k):=\frak{e}_{-1}(k)+\frak{e}_1(k).
$$
Note that
$$
\int_{\bbT}\frak{e}_1(k)dk=\int_{\bbT}\frak{e}_{-1}(k)dk=1
$$
and 
$$
R(k)=\frac{3}{4}\frak r(k)=2\sin^2(\pi k)\left[1+2\cos^2(\pi k) \right].
$$
The  process   $K_t(k)$ is a jump Markov process of the type considered in Section \ref{sec3.4}.  The  mean jump time and the transition probability operator of  the skeleton Markov chain $\{\xi_n,\,n\ge0\}$ are given by 
 $\theta(k)=R^{-1}(k)$
   and  
\begin{eqnarray*}
&&
 Pf(k):=\theta(k)\int_{\bbT}R(k,k')f(k')dk' \\
 &&
=\sum_{\iota\in\{-1,1\}}\langle \frak e_\iota,f\rangle \frac{\frak e_{-\iota}(k)}{\frak r(k)},\quad f\in C(\bbT),
 \end{eqnarray*}
 respectively.
Probability measure
 $
 \pi(dk)=(1/2)\frak r(k)dk
 $
 is reversible under the dynamics of the chain.  It is  clear that Condition \ref{kernel-a} holds.
 It has been shown in Section 3 of \cite{jko} that  Condition \ref{sg} is satisfied.

Let $\{Q_t,\,t\ge0\}$ be the semigroup corresponding to the generator ${\cal L}$. It can easily be argued that
$Q_t$ is a contraction on $L^p(\bbT)$ for any  $p\in[1,+\infty]$.
We shall need the following estimate.
 \begin{thm}
 \label{L1-bounds}
For a given $a\in(0,1]$ there exists $C>0$
\begin{equation}
\label{041302}
\|Q_{t}f\|_{L^1(\bbT)}\le  \frac{C}{(1+t)^a}\|f\|_{{\cal B}_a},\quad\forall \,t\ge0,\
\end{equation}
for all $f\in {\cal B}_a$ such that $\int_{\bbT} fdk=0$.
\end{thm}
The proof of this result shall be presented in Section \ref{sec6.2}. We proceed first with its application in the proof of Theorem \ref{prop1}.
The additive functional appearing in \eqref{bar-U} shall be denoted by
$$
Y^{(N)}_t(k):=\frac{1}{N^{2/3}} \int_0^{Nt}\om'(K_s(k))ds,
$$
or by $Y^{(N)}_t$ in case it corresponds to the stationary process $K_t$.
It is of the type considered in Section \ref{jump-process}
with $\alpha=3/2$ and 
 \begin{equation}
 \label{Psi}
 \Psi(k):=\om'(k)\theta(k).
 \end{equation}
Since the dispersion relation satisfies
 \begin{equation}
 \label{011505}
 \om(k)=|k|\left[\frac{\hat\al''(0)}{2}+O(k^2)\right]^{1/2}\quad\mbox{for }k\ll1,
 \end{equation}
 we have
 $$
 \om'(k)={\rm sgn\, }k\left[\frac{\hat\al''(0)}{2}+O(k^2)\right]^{1/2}\quad\mbox{for }k\ll1
 $$
and
$$
 \left|\pi(\Psi>\la)-\frac{c_*^+}{\la^{3/2}}\right|\le \frac{C^*}{\la^2},\quad \forall\,\la\ge1,
 $$
   with 
   $$
   c_*^+:=2^{-1/4}3^{-5/2}\pi^{1/2}[\hat\al''(0)]^{3/4} 
   $$
   and some $C^*>0$. 
     Condition \ref{tail} is therefore satisfied with $\alpha=3/2$ and arbitrary
  $\alpha_1<1/2$.  Since $\Psi(k)$ is odd  we have $c_*^-=c_*^+$.   On the other hand, jump mean time
$\theta(k)$ satisfies \eqref{theta} with $\al_2=3/2$.

%

We  can apply to  $Y^{(N)}_t$ the conclusion of part 1) of Theorem \ref{main}. In this case $\tilde c_*(\la)\equiv \hat c$, where
  $$
  \hat c=\frac{3}{2^{1/2}}\bar\theta^{-3/2}\Gamma\left(\frac52\right)c_*^+\int_0^{+\infty}\frac{\sin^2 x}{x^{5/2}}dx.
  $$  
  Since the integral on the right hand side equals $4\sqrt{\pi}/3$ 
  we obtain \eqref{031207}. 
Let $Z_t$ be the corresponding  symmetric $3/2$-stable process. From  \eqref{bar-W} we obtain
$\overline W(t,p)=\overline W(p)\bbE e^{-ip Z_t}.$ Therefore,
for any $J(\cdot)\in{\cal S}$
\begin{eqnarray}
\label{130607}
&&|\langle W_N(t),J\rangle-\langle \overline W(t),J\rangle|\\
 && =\left| \int_{\bbR\times \bbT} J^*(p,k)\bbE\left[ W(p,K_{Nt}(k))e^{- i p Y^{(N)}_t(k)}- \overline W(p)e^{- i p Z_t}\right]dp dk\right|.\nonumber
\end{eqnarray}

Let $1>\beta>1/3$. The left hand side of \eqref{130607} is estimated by $E_1+E_2+E_3$, where
\begin{eqnarray*}
&& E_1:=\left| \int_{\bbR\times \bbT} J^*(p,k)\bbE\left\{  W(p,K_{Nt}(k))\left[e^{- i p Y^{(N)}_t(k)}- e^{- i p Y^{(N)}_{t(1-N^{-\beta})}(k)}\right]\right\}dpdk\right|,\\
&&
E_2:=\left| \int_{\bbR\times \bbT} J^*(p,k)\bbE \widetilde W_{p}(K_{Nt}(k))e^{-i  p Y^{(N)}_{t(1-N^{-\beta})}(k)}dpdk\right|,\\
&&
E_3:=\left| \int_{\bbR\times \bbT} J^*(p,k)\overline W(p)\bbE \left[e^{- i p Y^{(N)}_{t(1-N^{-\beta})}(k)}- e^{- i p Z_t}\right]dpdk\right|
\end{eqnarray*}
and $\widetilde W_{p}(k):=  W(p,k)-\overline W(p) $.
The first term can be estimated as follows
\begin{eqnarray}
\label{021307}
&& E_1\le\|\om'\|_{\infty}tN^{1/3-\beta} \int_{\bbR\times \bbT} |p| |J(p,k)|\bbE |W(p,K_{Nt}(k))|dpdk\nonumber
\\
&&
\le CtN^{1/3-\beta} \|J\|_{{\cal A}_1'}\| W\|_{\cal A}.
\end{eqnarray}

To estimate the second term note that by the Markov property we can write
\begin{eqnarray}
\label{010707}
&& \bbE\left[  \widetilde W_{p}(K_{Nt}(k))e^{- i p Y^{(N)}_{t(1-N^{-\beta})}(k)}\right]\\
&&
=\bbE\left[  Q_{N^{1-\beta}t} \widetilde W_{p}(K_{Nt(1-N^{-\beta})}(k))e^{- i p Y^{(N)}_{t(1-N^{-\beta})}(k)}\right].\nonumber
\end{eqnarray}
Term $E_2$ can be therefore estimated by
$$
\|J\|_{{\cal A}'}\sup_{p\in\bbR}\|Q_{N^{1-\beta}t} \widetilde W_{p}\|_{L^1(\tilde\pi)}.
$$
 Invoking Theorem \ref{L1-bounds}  we obtain
$$
\sup_{p\in\bbR}\|Q_{N^{1-\beta}t} \widetilde W_{p}\|_{L^1(\bbT)}\le  \frac{C}{N^{a(1-\beta)}}\|W\|_{{\cal B}_a},\quad\forall \,t\ge 1,\,N\ge1.
$$
As a result we conclude immediately that
\begin{equation}
\label{041307}
E_2\le \frac{C}{N^{a(1-\beta)}}\|W\|_{{\cal B}_a}\|J\|_{{\cal A}'}.
\end{equation}

To deal with $E_3$ 
note that by reversibility of the process $K_t$ it equals

$$
E_3=\left| \int_{\bbR} \overline W(p)\bbE \left[J^*(p,K_{t(1-N^{-\beta})})e^{- i p  Y^{(N)}_{t(1-N^{-\beta})}}-\bar J^*(p) e^{- i p Z_t}\right]dp\right|
$$
where  $\bar J(p):=\int_\bbT J(p,k)dk$. We obtain that
\begin{eqnarray*}
&&
E_3
\le \left| \int_{\bbR}\overline W(p)\bbE\left\{  J^*(p,K_{t(1-N^{-\beta})})\left[e^{- i p Y^{(N)}_{t(1-N^{-\beta})}}-e^{- i pY^{(N)}_{t(1-2N^{-\beta})}}\right]\right\}dp\right|\\
&&
+\left| \int_{\bbR} \overline W(p)\bbE \left[\tilde J^*_p(K_{t(1-N^{-\beta})}) e^{- i p Y^{(N)}_{t(1-2N^{-\beta})}}\right]dp\right|\\
&&
+\left| \int_{\bbR} \bar J^*(p)\overline W(p)\bbE \left[e^{- i p Y^{(N)}_{t(1-2N^{-\beta})}}- e^{- i p Z_t}\right]dp\right|\\
&&
=E_{31}+E_{32}+E_{33}.
\end{eqnarray*}
From this point on  handle this term similarly to  what has been  done before and obtain that
\begin{equation}
\label{021307a}
 E_{31}\le \frac{Ct}{N^{\beta-1/3}} \|J\|_{{\cal A}_1'}\| W\|_{\cal A}
\end{equation}
and
\begin{equation}
\label{041307a}
E_{32}\le \frac{C}{N^{(1-\beta)a}}\|W\|_{{\cal A}}\|J\|_{{\cal B}_{a,b}},
\end{equation}
provided $b>1$.

Term $E_{33}$ can be handled with the help of  Theorem \ref{main} with $\al=\al_2=3/2$ and an arbitrary $\al_1<1/2$. Therefore, for any $\delta<2/13$  we can find  a constant $C>0$ such that
\begin{eqnarray}
\label{031307}
&& E_{33}\le \frac{C}{N^{\delta}}(t+1)\| W\|_{{\cal A}} \|J\|_{{\cal A}_{5}'}.
\end{eqnarray}
We have reached therefore the conclusion of Theorem \ref{prop1} with the exponent 
$\ga'$
as indicated in \eqref{011005}.

\qed

\subsection{Proof of Theorem \ref{prop1a}}
\label{sec6.3}
The respective additive functional in this case equals
$$
Y^{(N)}_t(k):=\frac{1}{N^{1/2}} \int_0^{Nt}\om'(K_s(k))ds,
$$
where $N:=\eps^{-2\ga}$. The observable $\Psi$ is given by \eqref{Psi}.
 From \eqref{011505} we get
 $$
\om'(k)=\frac{\hat\al''(0)k}{2}\left[\hat\al(0)+\frac{\hat\al''(0)}{2}k^2\left(1+O(k^2)\right)\right]^{-1/2}\quad\mbox{for }k\ll1.
 $$
 and the asymptotics of the tails of $\Psi(k)$ is given by 
  $$
 \pi(|\Psi|>\la)\le \frac{C^*}{\la^{3}} 
 $$ for
   some $C^*>0$ and all $\la\ge1$. The observable belongs therefore to $L^2(\pi)$ and, since it is odd,   its mean is $0$. 
     Conditions \ref{taila} and \ref{lower-t} are therefore fulfilled.
The latter with $\al_2=3$.    We also have
   \begin{equation}
   \label{031207a}
\hat c:= 9\si^2,
 \end{equation}
where
\begin{equation}
   \label{031207b}
 \si^2=\int_{\bbT}[\chi^2(k)-(P\chi)^2(k)
 ]\pi(dk)
 \end{equation}
and $\chi$ is the unique zero mean solution of  equation 
\begin{equation}
   \label{031207c}\chi-P\chi=\Psi.
    \end{equation}
The result is then the consequence of the argument made in Section \ref{sec6.1} and Theorems \ref{main} and  \ref{L1-bounds}

\subsection{Proof of Theorem \ref{L1-bounds}}

\label{sec6.2}

Denoting $f_t:=Q_tf$ we can write, using Duhamel's formula 
\begin{equation}
\label{duhamel}
f_t=S_tf+\frac{3}{4}\sum_{\iota\in\{-1,1\}} \int_0^tS_{t-s}\frak e_{-\iota}(k)\langle \frak e_\iota,f_s\rangle ds,
\end{equation}
where
$$
S_tf(k):=e^{-R(k)t}f(k).
$$
Let   $
\mathbb H:=[\la\in\mathbb C:\,{\rm Re\,}\la>0], 
 $
$$
\hat f(\lambda):=(\la-{\cal L})^{-1}f=\int_0^{+\infty}e^{-\la s}f_s ds,\quad \la\in\mathbb H,
$$
\begin{equation}
\label{021505}
\hat{\frak f}_{0}(\la):=\frac{f}{\frak{r}+\la}
\end{equation}
and 
$\hat{\frak f}_{\iota}(\la):=\langle \hat f(\lambda),\frak e_\iota\rangle$, $\iota=\pm1$.
Formula \eqref{021505} extends to  the resolvent set of the generator ${\cal L}$ in $L^2(\bbT)$, that contains in particular
$\mathbb C\setminus[-M,0]$, with  $M:=(4/3)\|R\|_\infty+1$. 

From \eqref{duhamel} we obtain that 
\begin{equation}
\label{laplace}
\hat f(\lambda)=\frac43 \hat{\frak f}_{0}\left(\frac{4\la}{3}\right)+\sum_{\iota\in\{-1,1\}}\hat{\frak f}_{\iota}\left(\la\right)\frac{\frak e_{-\iota}}{\frak{r}+4\la/3}.
\end{equation}
Vector $\hat{ \frak f}^{\tiny T}(\lambda):=[\hat{ \frak f}_{-1}(\lambda),\hat{ \frak f}_{1}(\lambda)]$
 satisfies therefore
 \begin{equation}
 \label{031005}
 \hat{ \frak f}(\lambda)={\frak A}^{-1}\left(\frac{4\la}{3}\right)\hat{ \frak g}\left(\frac{4\la}{3}\right),
\end{equation}
 where  
$$
 {\frak A}(\la)=\left[
 \begin{array}{lll}
 a(\la)&&a_{-1}(\la)\\
 &&\\
 a_{1}(\la)&& a(\la)
 \end{array}\right],
$$
$$
a(\la):=1-\int_{\bbT}\frac{\frak e_{-1}(k) \frak e_{1}(k)}{\la+\frak{r}(k)}dk,
$$
$$
a_\iota(\la):=-\int_{\bbT}\frac{\frak e_{\iota}^2(k) dk}{\la+\frak{r}(k)},
$$
 $\hat{ \frak g}^{\tiny T}(\lambda):=[\hat{ \frak g}_{-1}(\lambda),\hat{ \frak g}_{1}(\lambda)]$
 and
 $$
 \hat{ \frak g}_{\iota}(\lambda)=\frac43\int_{\bbT}\frac{f(k) \frak e_\iota(k)dk}{\la+\frak{r}(k)},\quad\iota=\pm1,\,\la\in\mathbb C\setminus[-M,0].
 $$
 Let $
 \Delta(\la):={\rm det\,}{\frak A}(\la)
 $
 and
 $$
b_\iota(\la):=-\int_{\bbT}\frac{\frak e_{\iota}(k) dk}{\la+\frak{r}(k)}.
$$
 Observe that
 \begin{equation}
 \label{021005}
 a(0)=-a_{-1}(0)=-a_{1}(0)
 \end{equation}
 and
 \begin{equation}
 \label{011205}
  \Delta(\la)=\la[\la b_{-1}(\la)b_1(\la)+b_{-1}(\la)a_1(\la)+a_{-1}(\la)b_1(\la)].
 \end{equation}
From \eqref{021005} we get that $\Delta(0)=0$.  
In addition, 
from \eqref{011205} 
we can see that $D(\la):= \Delta(\la)\la^{-1}$ is analytic in $\mathbb C\setminus[-M,0]$. 
In addition,
$$
\lim_{\la\to0,\la\in\overline{\mathbb H}}D(\la)=b_{-1}(0)a_1(0)+a_{-1}(0)b_1(0)>0.
$$
Hence, there exist $\varrho>0$ and $c_*>0$ such that
 \begin{equation}
 \label{051005}
|\Delta(\la)|\ge c_*|\la|,\quad \forall\,\la\in\overline{\mathbb H},\,|\la|\le \varrho.
 \end{equation}
 It can be straightforwardly argued that  $\varrho$ can be further adjusted in such a way that
\eqref{051005} holds on the boundary ${\cal C}$ of the rectangle $(-M,0)\times (-\varrho,\varrho)$.

 From \eqref{031005} we obtain that
 \begin{equation}
 \label{011105}
 \hat{ \frak f}_{\iota}(\lambda)=\Delta^{-1}\left(\frac{4\la}{3}\right)\frak{n}_{\iota}\left(\frac{4\la}{3}\right),\quad\iota=\pm1,
 \end{equation}
 where
 $$
\frak{n}_{-1}(\la) :=a(\la)\hat{ \frak g}_{-1}(\la)- a_{-1}(\la)\hat{ \frak g}_{1}(\la)
$$
 and
  $$
\frak{n}_1(\lambda):= -a_{1}(\la)\hat{ \frak g}_{-1}(\la)+a(\la)\hat{ \frak g}_{1}(\la).
 $$
 Using the fact that $\int_{\bbT}f dk=0$, after a straightforward calculation, we obtain  that
 \begin{equation}
 \label{021205}
 \frak{n}_{\iota}(\la)=\la \left[a_{\iota}(\la)\hat{\frak{g}}_0(\la)-b_{\iota}(\la)\hat{ \frak g}_{\iota}(\la)\right],\quad\iota=\pm1,
 \end{equation}
where
$$
\hat{\frak{g}}_0(\la):=\int_{\bbT}\hat{\frak{f}}_0(\la,k)dk.
$$
%
%
%
%
%
%
%
%
%
%
%
%
%
%
%

 Using the well known formula, see e.g. Chapter VII.3.6 of \cite{dunford-schwartz},
 $$
 Q_t f=\frac{1}{2\pi i}\int_{{\cal K}}e^{\la t}(\la-{\cal L})^{-1}fd\la,
 $$
 where ${\cal K}$ is a  contour  enclosing the 
$L^2$  spectrum of ${\cal L}$, that as we recall is contained in $[-M,0]$. It is easy to see that
\begin{equation}
\label{041505}
|\hat{ \frak f}_{\iota}(\lambda)|\le C\|f\|_{{\cal B}_a}|\la|^{a-1},\quad \la\in \overline{\mathbb H},\,|\la|\le \varrho,\,\iota=0,\pm1
\end{equation}
for an appropriate $\varrho>0$. This is clear for  $\iota=0$.
Form this and   \eqref{051005},  \eqref{021205} we conclude that \eqref{041505} holds also for $\iota=\pm1$. 
%
%
%
%
%
%
%
%
%
%
%
%
Therefore, we can use as ${\cal K}$  the boundary ${\cal C}$ of the rectangle, mentioned after  \eqref{051005},  oriented counter-clockwise.
   From \eqref{laplace}  we get
 \begin{equation}
\label{duhamel1}
f_t=S_tf+\frac{1}{2\pi i}\sum_{\iota\in\{-1,1\}}\frak e_{-\iota}\int_{{\cal C}}\frac{e^{\la t}\frak{n}_{\iota}\left(4\la/3\right)d\la}{(\frak{r}+4\la/3)\Delta\left(4\la/3\right)}.
\end{equation}
Note that
$$
\|S_tf\|_{L^1(\bbT)}=\int_{\bbT}e^{-R(k)t}|f(k)|dk\le \frac{C}{t^a}\|f\|_{{\cal B}_a},
$$
where $C:=\sup_{x\ge 0}x^ae^{-x}$. Consider the term $I(t)$ corresponding to the integral on the right hand side of \eqref{duhamel1}. We can write $I(t)=\sum_{i=1}^4I_i(t)$, where $I_i(t)$ correspond to the sides of the rectangle  $\{-M\}\times (-\varrho,\varrho)$,  $[-M,0]\times \{-\varrho\}$,  $[-M,0]\times \{\varrho\}$, $\{0\}\times (-\varrho,\varrho)$ appropriately oriented. The estimations of $I_i(t)$, $i=1,2,3$ are quite straightforward and lead to the bounds
\begin{equation}
\label{i3}
\|I_i(t)\|_{L^1(\bbT)}\le \frac{C}{t+1}\|f\|_{L^1(\bbT)},\quad i=1,2,3.
\end{equation}
%
%
%
%
%
%
%
%
%
%
%
%
%
%
%
%
%
%
%
%
%
%
%
%
%
%
%
%
%
%
%
%
To deal with $I_4(t)$ observe that, thanks to \eqref{021205}, it equals to $I_{41}(t)+I_{42}(t)$, where
$$
I_{4,j}(t):=\frac{1}{2\pi i}\sum_{\iota\in\{-1,1\}}\frak{e}_{-\iota}\int_{-\varrho}^{\varrho}e^{i\nu t}(g_{\iota,j}D^{-1})\left(\frac{4i\nu}{3}\right)d\nu,\quad j=1,2
$$
and
$$
g_{\iota,1}(\nu)=\frac{a_{\iota}\left(i\nu\right)\hat{\frak{g}}_{0}\left(i\nu\right)}{\frak{r}+i\nu},\qquad g_{\iota,2}(\nu)=\frac{-b_{\iota}\left(i\nu\right)\hat{\frak g}_{\iota}\left(i\nu\right)}{\frak{r}+i\nu}.
$$
The asymptotics of $I_{4,1}$ for $t\gg1$ is, up to a term of order $\|f\|_{L^1(\bbT)}/t$, the same as
$$
\tilde I_{4,1}(t):=\frac{1}{2\pi i}\sum_{\iota\in\{-1,1\}}\frak{e}_{-\iota}\int_{\bbR}e^{i\nu t}(Fg_{\iota,1})\left(4i\nu/3\right)d\nu
$$
for some $C^\infty$ function $F(i\nu)$ supported in $(-\rho,\rho)$ and equal to $D^{-1}(\i\nu)$ in  $(-\rho/2,\rho/2)$.
Denoting
$$
{\frak h}(\la):=\frac{\hat{\frak g}_0\left(\la\right)}{{\frak r}+\la}
$$
we can write
\begin{eqnarray*}
&&
\|\tilde I_{4,1}(t)\|_{L^1(\bbT)}\le \frac{1}{4\pi }\sum_{\iota\in\{-1,1\}}\int_{\bbT}\frak{e}_{-\iota}dk\\
&&
\times\left|
\int_{\bbR}e^{i\nu t}\left[(Fg_{\iota,1})\left(\frac{4i}{3}\left(\nu+\frac{\pi}{t}\right)\right)-(Fg_{\iota,1})\left(\frac{4i\nu}{3}\right)\right]d\nu\right|\\
&&
\le \frac{1}{4\pi }\sum_{\iota\in\{-1,1\}}\int_{\bbT}\frak{e}_{-\iota}dk\int_{-2\varrho}^{2\varrho}\left|{\frak h}\left(4i(\nu+\pi/t)/3\right)\left[(Fa_{\iota})\left(\frac{4i}{3}\left(\nu+\frac{\pi}{t}\right)\right)-(Fa_{\iota})\left(\frac{4i\nu}{3}\right)\right]\right|d\nu\\
&&
+\frac{1}{4\pi }\sum_{\iota\in\{-1,1\}}\int_{\bbT}\frak{e}_{-\iota}dk\int_{-2\varrho}^{2\varrho}\left|(Fa_{\iota})\left(\frac{4i\nu}{3}\right)\left[{\frak h}\left(4i\nu/3\right)-{\frak h}\left(4i(\nu+\pi/t)/3\right)\right]\right|d\nu
\end{eqnarray*}
for sufficiently large $t$ (the support of $F$ is contained in $(-\varrho,\varrho)$).
The first term on the utmost right hand side can be estimated by
$$
\frac{C}{t}\int_{-2\varrho}^{2\varrho}\left|\hat{\frak g}_0\left(4i(\nu+\pi/t)/3\right)\right|d\nu\le 
\frac{C}{t}\int_{-2\varrho}^{2\varrho}\int_{\bbT}\frac{|f|d\nu dk}{{\frak r}^a|\nu+\pi/t|^{1-a}}dk\le \frac{C}{t}\|f\|_{{\cal B}_a}.
$$
As for the second term, it can be estimated by
\begin{eqnarray}
\label{011405}
&&
\frac{C}{t}\sum_{\iota\in\{-1,1\}}\int_{\bbT}\int_{-2\rho}^{2\rho}\frac{\frak{e}_{-\iota}|\hat{\frak{g}}_{0}\left(4i(\nu+\pi/t)/3\right)|dkd\nu}{|\frak{r}+4i\nu/3||\frak{r}+4i(\nu+\pi/t)/3|}\\
&&
+\frac{C}{t}\sum_{\iota\in\{-1,1\}}\int_{\bbT}\int_{-2\rho}^{2\rho}\frac{\frak{e}_{-\iota}dkd\nu}{|\frak{r}+4i\nu/3|}\int_{\bbT}\frac{|f|dk}{|\frak{r}+4i\nu/3||\frak{r}+4i(\nu+\pi/t)/3|}.\nonumber
\end{eqnarray}
The first term is less than
\begin{equation}
\label{021405}
\frac{C}{t}\sum_{\iota\in\{-1,1\}}\int_{\bbT}\frac{\frak{e}_{-\iota}dk}{\frak{r}^{1+b}}\int_{\bbT}\frac{|f|dk}{\frak{r}^a}
\int_{-2\rho}^{2\rho}\frac{ d\nu}{|\nu|^{1-b}|\nu+\pi/t|^{1-a}}
\end{equation}
for some $b\in(0,1/2)$. Since for any $a,b>0$ there exists $C>0$ such that
\begin{equation}
\label{021405a}
\int_{-2\rho}^{2\rho}\frac{ d\nu}{|\nu|^{1-b}|\nu+x|^{1-a}}\le \frac{C}{x^{1-a-b}},\quad\forall\,x>0
\end{equation}
expression in \eqref{021405} can be estimated by $C\|f\|_{{\cal B}_a}/t^{a+b}$.
Finally the second term in \eqref{011405} can be estimated by
$$
\frac{C}{t}\int_{\bbT}\int_{-2\rho}^{2\rho}\frac{d\nu}{|\nu|^{1-a/2}|\nu+\pi/t|^{1-a/2}}\int_{\bbT}\frac{|f|dk}{\frak{r}^a}\le \frac{C}{t^a}\|f\|_{{\cal B}_a},
$$
by virtue of \eqref{021405a}.
 Summarizing, we have shown that
$$
\|\tilde I_{4,1}(t)\|_{L^1(\bbT)}\le \frac{C}{(t+1)^a}\|f\|_{{\cal B}_a}
$$
for some $C>0$ and all $t>0$. The estimates for $\|\tilde I_{4,2}(t)\|_{L^1(\bbT)}$ are quite analogous.
 
 {\bf Acknowledgements.} Both authors wish to thank the anonymous  referee of the paper for many valuable remarks that contributed to the improvement of the manuscript. T. K. and L. S. acknowledge the support of the grant of the Polish Ministry of Science and Higher Education, NN 201 419 139.

\end{document}